\documentclass[journal]{IEEEtran} 
\usepackage{cite}
\ifCLASSINFOpdf
  \usepackage[aboveskip=10pt,belowskip=-15pt,font=footnotesize]{caption}
\else
  \usepackage[dvips]{graphicx}
\fi
\usepackage{amssymb,bm,mathrsfs,bbm,amscd,amsmath}
\newtheorem{Proposition}{Proposition}
\newtheorem{Remark}{Remark}
\newtheorem{Lemma}{Lemma}
\usepackage{paralist}
\usepackage{subfig}
\usepackage{tabularx}
\usepackage{multirow}
\usepackage[ruled]{algorithm2e}  

\SetKwInput{KwIn}{\textbf{Input}}
\SetKwInput{KwData}{\textbf{Initialization}}
\SetKwInput{KwOut}{\textbf{Output}}
\usepackage{verbatim}
\usepackage[bookmarks,colorlinks]{hyperref}

\usepackage{color}

    \def\Complex{{\rm\rule[.23ex]{.03em}{1.1ex}\kern-.3em{C}}}

    \newcommand{\be}{\begin{equation}} \newcommand{\ee}{\end{equation}}
    \newcommand{\bea}{\begin{eqnarray}} \newcommand{\eea}{\end{eqnarray}}
    \newcommand{\benum}{\begin{enumerate}} \newcommand{\eenum}{\end{enumerate}}

        \newcommand{\qa}{\mathbf{a}}

        \newcommand{\qh}{\mathbf{h}}

        \newcommand{\qm}{\mathbf{m}}
        \newcommand{\qn}{\mathbf{n}}

        \newcommand{\qv}{\mathbf{v}}
        
        \newcommand{\qx}{\mathbf{x}}
        \newcommand{\qy}{\mathbf{y}}

        \newcommand{\qA}{\mathbf{A}}

        \newcommand{\qD}{\mathbf{D}}

        \newcommand{\qH}{\mathbf{H}}
        \newcommand{\qI}{\mathbf{I}}

        \newcommand{\qzero}{\mathbf{0}}

        \newcommand{\qSigma}{\boldsymbol{\Sigma}}

        \newcommand{\qalpha}{{\boldsymbol \alpha}}
        \newcommand{\qbeta}{{\boldsymbol \beta}}

        \newcommand{\qgamma}{\boldsymbol{\gamma}}

        \newcommand{\qmu}{{\boldsymbol \mu}}

        \newcommand{\calA}{{\mathcal A}}
        
        \newcommand{\calC}{\mathcal{C}}

        \newcommand{\calF}{{\mathcal F}}
        
        \newcommand{\calK}{\mathcal{K}}

        \newcommand{\calX}{\mathcal{X}}
        
        \newcommand{\calCN}{\mathcal{CN}}

        \newcommand{\diag}{{\sf diag}}

        \newcommand{\tr}{{\sf tr}}
        \newcommand{\Ex}{{\sf E}}
        
        \newcommand{\Varx}{{\sf Var}}
        \newcommand{\mse}{{\sf mse}}
        \newcommand{\KL}{{\sf KL}}

        \newcommand{\leftvec}{\buildrel{\lower3pt\hbox{$\scriptscriptstyle\leftarrow$}}\over}
        \newcommand{\rightvec}{\buildrel{\lower3pt\hbox{$\scriptscriptstyle\rightarrow$}}\over}
        
        \newcommand{\lqx}{\leftvec{\qx}}
        
        \newcommand{\rqx}{\rightvec{\qx}}

        \newcommand{\bl}[1]{\color{blue}#1}

\begin{document}
%
\title{Expectation Propagation Detector for\\
Extra-Large Scale Massive MIMO}
%
%
%

\author{Hanqing~Wang,~
	    Alva~Kosasih,~
        Chao-Kai~Wen,~
        Shi~Jin~
        and~Wibowo~Hardjawana~
\thanks{H. Wang and S. Jin are with the National Mobile Communications Research Laboratory, Southeast University, Nanjing 210096, P. R. China. P (e-mail: $\rm hqwanglyt@seu.edu.cn;~jinshi@seu.edu.cn$).}
\thanks{C.-K. Wen is with the Institute of Communications Engineering, National Sun Yat-sen University, Kaohsiung 804, Taiwan (e-mail: $\rm chaokai.wen@mail.nsysu.edu.tw$).}
\thanks{A. Kosasih and W. Hardjawana are with the Wireless Engineering Laboratory, University of Sydney, Sydney, Australia (e-mail: $\rm alva.kosasih@sydney.edu.au;~wibowo.hardjawana@sydney.edu.au$).}

}

\maketitle

\begin{abstract}

The order-of-magnitude increase in the dimension of antenna arrays, which forms extra-large-scale massive multiple-input-multiple-output (MIMO) systems, enables substantial improvement in spectral efficiency, energy efficiency, and spatial resolution. However, practical challenges, such as excessive computational complexity and excess of baseband data to be transferred and processed, prohibit the use of centralized processing. A promising solution is to distribute baseband data from disjoint subsets of antennas into parallel processing procedures coordinated by a central processing unit. This solution is called subarray-based architecture. In this work, we extend the application of expectation propagation (EP) principle, which effectively balances performance and practical feasibility in conventional centralized MIMO detector design, to fit the subarray-based architecture. Analytical results confirm the convergence of the proposed iterative procedure and that the proposed detector asymptotically approximates Bayesian optimal performance under certain conditions. The proposed subarray-based EP detector is reduced to centralized EP detector when only one subarray exists. In addition, we propose additional strategies for further reducing the complexity and overhead of the information exchange between parallel subarrays and the central processing unit to facilitate the practical implementation of the proposed detector. Simulation results demonstrate that the proposed detector achieves numerical stability within few iterations and outperforms its counterparts.
\end{abstract}


%
\IEEEpeerreviewmaketitle

\section{Introduction}
Massive multiple-input multiple-output (MIMO) is widely believed to be one of the key techniques for fifth-generation wireless systems \cite{Andrews-2014JSAC}.
This technique was proposed to equip a base station (BS) with a large-scale antenna array (in the order of hundreds or thousands) and simultaneously serve a relatively small number of mobile users in the same time-frequency resource \cite{Marzetta-2010TWC}.
Many advantages, including substantial improvement in spectral efficiency, energy efficiency, and spatial resolution, can be achieved by increasing the antenna array dimension \cite{Larsson-2014MCOM}.
This approach is expected to facilitate ultra high data rate and system throughput.
These advantages can be obtained by using a simple linear transceiver \cite{Lu-2014JSTSP}.
Massive MIMO communication systems have recently been implemented in the academia and various industries, and many prototyping systems have been constructed \cite{Yang-2017ChinaCom, Gao-2018ACCESS}, which implies the potential commercial success of these communication systems in the future.

Current cellular networks commonly deploy compact and co-located antenna array with small antenna separation, such as in the order of the wavelength, on the top of a tower or roof. 
However, the conventional compact antenna deployment encounters numerous practical challenges when the array dimension is increased to the order of thousands and more, such as the size and weight of the array and wind load. 
Moreover, to achieve great spatial resolution, it is desirable to distribute antennas over a substantially large area \cite{Bjornson2017}. 
These factors advocate embracing new antenna deployment strategies for arrays with extremely large dimensions.
One potential approach is to integrate the antenna array into large structures, such as along the walls of tall buildings, airports, or large shopping malls, or along the structure of a stadium \cite{Martinez-2014GCW}.
Another approach is to distribute disjoint subsets of antennas of the entire array over a large geographical area coordinated by a central processing unit, thus yielding distributed antenna systems \cite{Rheath-2013ACSSC}.
We refer to communication paradigms that utilize antenna arrays with the order-of-magnitude increase in their dimensions as extra-large-scale massive MIMO systems.

In extra-large-scale massive MIMO systems, the conventional processing architecture, in which signals received by all antennas and the full channel matrix are involved in the centralized processing, faces numerous challenges, especially in crowded scenarios.
An evident challenge is the excessive computation complexity due to the large number of antennas and users.
Even simple linear transceivers involve complex matrix operations to high-dimensional matrices.
Another prominent challenge is the need to transfer excessively large amounts of baseband data received by extra-large-scale antenna array to the baseband processing unit \cite{Larsson-2018TSP}.
A promising solution is to divide the entire antenna array into a few disjoint units, referred to as subarrays.
Each subarray is associated with an individual processing unit that accesses only its local signals \cite{Amiri-2019Arxiv} and performs parallel processing with reduced complexity to produce coarse estimate of the transmitted signal.
A central processing unit is responsible for producing refined estimate by performing a certain combination to the outputs of parallel subarrays.
This subarray-based architecture enables the use of parallel computation supporting components (e.g., graphics processing unit) and relaxes the bandwidth requirement of the interface circuits by deploying parallel interconnections between subarrays and their local processing units.
In practical systems, subarrays may correspond to separate hardware entities or software-defined logical interconnections between different portions of a signal array and their flexibly assigned processing resource \cite{Amiri-2018GCW}. 
The number and size of subarrays are fixed in the former, whereas they are adjustable in the latter. 

Spatial non-stationary channel property generally occurs on large-scale antenna arrays \cite{Amiri-2019Arxiv}.
The term ``non-stationary" means that the different parts of the array may observe the same channel paths with varying power or distinct channel paths \cite{Liu-2012WCM,Gao-2013ACSSC}.
Particularly, energy received from each user by different portions of the array varies.
In this case, majority of energy received a specific user concentrates on small portions of the entire array, thus, inherent sparsity is introduced in the spatial dimension of the channel matrix \cite{Amiri-2018GCW}.
Meanwhile,``spatial stationary" means that the entire array receives approximately the same amount of energy from each user, and it usually features MIMO channels where a moderate number of (several tens of) antennas are compactly deployed. 
When spatial non-stationarity occurs, this subarray-based processing architecture becomes favorable.
First, each subarray only needs to detect signals transmitted by users that have sufficiently large received energy on that subarray, thereby enabling complexity reduction. 
Moreover, spatial non-stationary channel properties complicates the modeling of the channel on the entire array, which harms accurate channel estimation.
Nevertheless, simplification of channel acquisition can be realized by decomposing the entire array into subarrays whose channels are approximated as stationary \cite{Martinez-2016PIMRC}, which inspires detector development based on subarrays. 

This study focuses on the development of a subarray-based multiuser detector for extra-large-scale massive MIMO systems.
A few preliminary studies have been conducted on this topic.
The performance of the linear receiver for such a scenario was evaluated in \cite{RHeath-2019WCL}.
The subsequent work \cite{Amiri-2018GCW} adopted the successive interference cancellation (SIC) principle, which detects a signal from a specific user from a subarray with favorable interference conditions and then removes its contributions from the other subarrays to achieve performance improvement. 
The main drawback of this work is that detection of a certain user is performed from one subarray only, therefore, the contributions from other subarrays to the signal detection of this user cannot be combined. 
Moreover, detection of different users is performed in a serial manner, which does not fully exploit the parallel computation potential of subarray-based architecture.
These drawbacks motivate the use of advanced techniques to pursue better design.
The researchers in \cite{Jespedes-2014TCOM} proposed a MIMO detector based on expectation propagation (EP) and conventional centralized processing architecture and reported that the proposed detector outperforms many classic and state-of-the-art solutions with moderate computational complexity.
An integrated circuit design for the detector proposed in \cite{Jespedes-2014TCOM} was presented in \cite{Tang-2018ISSCC}, which showed that EP-based design effectively balances performance and practical feasibility.		 
These facts inspire us to extend the application of EP to subarray-based processing architecture. 
The proposed detector is constructed based on the general form of the EP principle \cite{Minka-2005}, and several analytical results are derived.  
Moreover, additional strategies are proposed to reduce complexity and information exchange overhead, and improve parallelism, to facilitate the practical implementation of the proposed design.
We derive the following meaningful findings and insights, which could provide guidance to further studies.

\begin{itemize}
	\item \emph{Algorithm derivation}: The proposed detector is derived via message updating and passing based on EP principle \cite{Minka-2001,Minka-2005,Vehtari-2014Arxiv} in a factor graph that describes the subarray-based architecture.
	This derivation procedure reveals the underlying logic and rationale of the proposed algorithm and makes it more than a purely heuristic one.
	Many extant methods, such as \cite{Jespedes-2014TCOM,Yuan-2014TWC,JMA-2016SPL,JMa-2016Access}, can be derived using the same derivation framework, which implies that the EP principle will continue to help in developing effective algorithms by describing a particular application scenario with a factor graph on which feasible computation of messages is allowed. 
	
	\item \emph{Theoretical analyses}:
	Convergence of the proposed iterative procedure is intuitively verified through evolution analysis and is justified by simulation results.
	We prove that the asymptotic behavior of the proposed subarray-based EP detector matches the replica prediction of Bayesian optimal performance when the channel matrix has independent and identically distributed (i.i.d.) entries.
	The simulation results reveal that even with a non-i.i.d. channel matrix, the performance under different sizes and numbers of subarrays is similar to that under centralized EP detector \cite{Jespedes-2014TCOM}.
	This finding enables complexity minimization by including a small number of antennas in each subarray without substantially performance degradation.	
	 	
	\item \emph{Engineering efforts}: We also propose several strategies to facilitate efficient implementation of the proposed detector.
	Specifically, we demonstrate that the sparsity in the local channel matrix of each subarray derived from the non-stationarity allows for additional complexity reduction.
	Furthermore, a hierarchical implementation architecture is proposed to decompose the operations in each subarray into several parallel computations, thus improving parallelism.
	A modification of the proposed detector is presented.
    In such modification,  only one feedforward from each subarray to the central processing unit is required in the entire procedure, which remarkably decreases the information exchange overhead and latency.	
\end{itemize}

\emph{Notations:} In this paper, we use lowercase and uppercase boldface letters to represent vectors and matrices, respectively.
For vector $\qa$, $a_j$ denotes the $j$-th entry of $\mathbf{a}$; $\Re{\qa}$ and $\Im{\qa}$ denote its real and imaginary parts respectively; and $[\qa]_{\calA}$ denotes the vector comprised by the elements of $\qa$ indexed by $\calA$.
For matrix $\qA$, we indicate its conjugate transpose, trace and inverse by $\qA^H$, $\tr(\qA)$ and $\qA^{-1}$ respectively.
The operator $\odot$ denotes the element-wise (Hadamard) product between two vectors or matrices with identical sizes.
The notation $\qI_N$ is used to denote the $N\times N$ identity matrix.
For set $\calA$, $|\calA|$ denotes its cardinality.
For random variable $x$, $\Ex(x)$ and $\Varx(x)$ denote the expectation and variance of $x$ respectively, while $\Ex(x\mid y)$ and $\Varx(x\mid y)$ denote the expectation and variance of $x$ conditioned on $y$ respectively.
The distribution of a proper complex Gaussian random variable $z$ with mean $\mu$ and variance $\nu$ is expressed by
$z\sim\mathcal{CN}(z;\mu,\nu)=\frac{1}{\pi\nu}\exp{\left(|z-\mu|^2/{\nu}\right)}$.
Lastly, the notation $:=$ is used to define a symbol as the expression on its right side.

\section{System Model}

We consider the scenario with $K$ single-antenna users served by a large-scale array with $N \geq K$ antennas in the same time frequency resource.
Specifically, we study the multi-user detector design.
In the considered scenario, each user encodes its own information bit stream and modulates it to be a sequence of constellation points in $\calX$ (e.g., 16-QAM).
For a certain time slot, we denote the transmit symbol of the $k$-th user by $x_k$, and stack the transmit symbols of $K$ users by the vector $\qx=[x_1,x_2,\ldots,x_K]^T$, where $\qx\in\calX^{K}$.
The average energy of the transmit symbols of different users is assumed to be the same and  denoted by $E_x$.
The narrowband\footnote{Although the detector is proposed for narrowband channel, its use in wideband channel can be supported by utilizing orthogonal frequency division multiplexing (OFDM) technique, which decomposes the convolutive multipath channels into a bank of parallel flat-fading subchannels.} equivalent baseband input-output relationship is given by
\begin{equation} \label{chan_model}
\qy = \qH \qx + \qn,
\end{equation}
where $\qy \in \mathbb{C}^N$ denotes the signal vector received by the antenna array,
$\qH \in \mathbb{C}^{N\times K}$ is the channel matrix,
and $\qn \in \mathbb{C}^N$ represents the additive white Gaussian noise (AWGN) vector with distribution $\calCN(\mathbf{0},\sigma^2 \qI_N)$.
We aim to produce a reliable estimate of $\qx$ with the knowledge of received signals $\qy$ and the channel matrix $\qH$ for hard or soft decision.

A promising solution for dealing with the excessively large data load and computational complexity caused by the order-of-magnitude increase in the antenna array dimension is to partition the antenna array into a certain number of subarrays, each associated with an independent processing entity.
Following this idea, we partition the whole array of $N$ antennas into $C \geq 1$ subarrays.
Let $N_c$ denote the number of antennas in subarray $c\in\calC=\{1,2,\ldots,C\}$, and thus $N = \sum_{c=1}^{C} N_c$.
Correspondingly, the received and AWGN vectors and channel matrix can be partitioned as $\qy = [\qy_1^T, \qy_2^T, \ldots, \qy_C^T]^T$, $\qH = [\qH_1^T,\qH_2^T,\ldots,\qH_C^T]^T$, and $\qn = [\qn_1^T,\qn_2^T,  \ldots, \qn_C^T]^T$.
Then, the received vector corresponding to the $c$-th subarray is given by
\begin{equation}\label{dec_chan_model_ST}
	\qy_c = \qH_c \qx + \qn_c,
\end{equation}
where ${\qH_c \in \mathbb{C}^{N_c\times K}}$ denotes the local channel matrix between $K$ users and antennas of subarray $c$,
and ${\qn_c \in \mathbb{C}^{N_c}}$ is noise vector corresponding to the $c$-th subarray.

\begin{figure}[!t]
	\centering
	\includegraphics[scale=0.32]{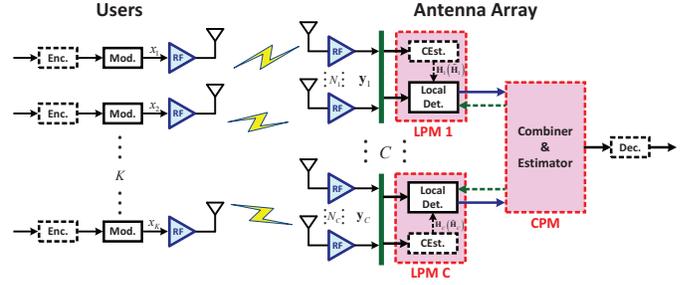}
	\caption{Block diagram of the subarray-based baseband processing architecture (Enc. = Encoder, Mod. = Modulator, CEst. = Channel Estimator, Det. = Detector and Dec. = Decoder).\label{F1}}
\end{figure}

We call \eqref{dec_chan_model_ST} the full subarray model in the subsequent sections because it involves full channel matrices.
The sparsity in the local channel matrices inherent in the spatial non-stationarity can be exploited to further reduce complexity.
When non-stationarity occurs, energy received from each user is concentrated on a small portion of subarrays, which is usually significantly smaller than $C$.
Specifically, only a limited portion of users can be viewed as being served by a specific subarray.
Therefore, the local channel matrices $\{\qH_c\}_{c=1}^{C}$ are approximately column-wise sparse, that is, the columns of $\qH_c$ corresponding the users with sufficiently large received energy on the $c$-th subarray have major values, while the other columns are close to zero.
The local channel matrix $\qH_c$ of each subarray can be estimated by applying compressive sensing techniques for multiple measurement vector problems (see \cite[Section 11.6]{Eldar2015} and references therein) and then be trimmed by removing these columns with values close to zero.
Accordingly, we denote the trimmed local channel matrix of subarray $c$ by $\tilde{\qH}_c$, and we define $\calK_c$ as the subset of index corresponding to the columns of $\qH_c$ which are retained in $\tilde{\qH}_c$. 
The matrix $\tilde{\qH}_c$ is a $N_c\times K_c$ submatrix of $\qH_c$ that comprises $K_c$ columns of $\qH_c$ indexed by $\calK_c$.
We then let $K_c=|\calK_c|$ and $\qx_c=[\qx]_{\calK_c}$.
Typically, $K_c$ is much smaller than $K$. 
With the above notations, the received model for each subarray is approximated by
\begin{equation}\label{dec_chan_model_NST}
\qy_c = \tilde{\qH}_c \qx_c + \qn_c,
\end{equation}
which is called as trimmed subarray model.
The reduction in the dimensions of local channel matrices will lead to reduced complexity in signal detection.

The subarray-based baseband processing architecture is shown in Fig. \ref{F1}.
Each subarray is associated with its local processing module (LPM), which accesses only their local  channel matrix and received vector.
In particular, the $c$-th subarray produces a coarse estimate of $\qx$ (or $\qx_c$) based on $\qy_c$ and $\qH_c$ (or $\tilde{\qH}_c$)\footnote{This paper mainly focuses on the detector design, thus we assume the local channel matrix, $\qH_c$ or $\tilde{\qH}_c$, is perfectly known at each subarray.
In practical systems, the local channel matrix of each subarray can be acquired by uplink pilot \cite{Lu-2014JSTSP}.}
which is feedforward to the central processing module (CPM).
The CPM is responsible for generating consensus information based on the information delivered by each subarray (solid blue lines), which is then broadcast back to all subarrays (dash green lines), and producing the final output, which is then applied to generate the input of the channel decoder.
The blue and green lines may correspond to dedicated data links, implemented by cables or optic fibers, or the data interfaces among chips or software modules depending on different architectures.
In the subsequent sections, we focus on detector design based on this subarray-based architecture.

\section{EP Detector for Subarray-based Architecture}
In this section, we exploit the EP principle to develop the detector for subarray-based architecture.
We consider the fundamental full subarray model \eqref{dec_chan_model_ST} to establish the technical foundation and facilitate a few theoretical analyses from the signal processing perspective.
We start by providing detailed interpretations of the proposed detector, followed by intuitive convergence justification and fixed point characterization.
Discussions related to efficient implementation of the proposed algorithm will be presented in the next section.

\subsection{Algorithm Derivation}
We develop a detector by constructing a reliable estimate of $\qx$ on the basis of the classical \emph{Bayesian inference framework} \cite{Kay1993}, which is initiated by calculating the a posteriori distribution as follows:
\begin{equation} \label{postPro}
p(\qx|\qy,\qH)   \propto\underbrace{\exp{\left(-\frac{\|\qy-\qH\qx\|^{2}}{\sigma^{2}}\right)}}_{\propto p(\qy|\qx,\qH)}\underbrace{\prod_{k=1}^{K} p(x_k)}_{=p(\qx)},
\end{equation}
where $p(\qy|\qx,\qH)$ denotes the likelihood function with the known channel matrix $\qH$ and $p(\qx)$ represents the a priori distribution of $\qx$.
Without loss of generality, we consider the case where $\{x_k\}_{k=1}^{K}$ are drawn independently from $\calX$ with equal probabilities, namely, $p(x_k=x) = 1/|\calX|$ for $\forall x \in \calX$.
We can then attain excellent inference in accordance with several optimal criteria.
For example, we can achieve the \emph{minimum mean-square error} (MMSE) criterion by computing the a posteriori distribution of $\qx$.
We can also achieve the \emph{maximum a posteriori} (MAP) criterion by searching $\qx$ over $\calX^N$ which maximizes the a posteriori distribution in \eqref{postPro}.
However, direct computation of the aforementioned estimates is intractable because of the calculation of the high-dimensional integral or exponential complexity.
This situation motivates us to pursue advanced mathematical tools to effectively approximate \eqref{postPro} with tractable complexity and appropriately fit the subarray-based architecture.
\begin{figure}[!ht]
	\centering
	\subfloat[Factor graph describing \eqref{dec_postPro}]{\includegraphics[scale=0.8]{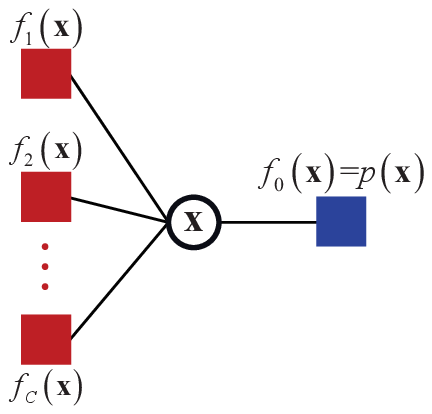}\label{F2A}}%
	\hspace{20pt}
	\subfloat[Factor graph describing \eqref{postPro}]{\includegraphics[scale=0.8]{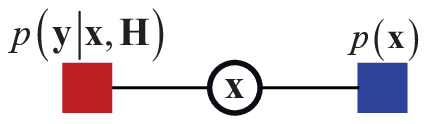}\label{F2B}}\\
	\caption{Factor graph used for algorithm derivation. The hollow circles
		represent the variable nodes and the solid squares represent the factor nodes.\label{F2}}
\end{figure}

EP is an effective tool that can be used to approximate complicated distributions, such as \eqref{postPro}, fitting appropriately the subarray-based architecture.
The basic idea is to represent \eqref{postPro} using a graphical model called factor graph associated with a certain factorization of \eqref{postPro}.
Compared with the conventional EP-based MIMO detector \cite{Jespedes-2014TCOM}, the factorization of \eqref{postPro} in this work further considers the subarray-based architecture \eqref{dec_chan_model_ST}, which is given by
\begin{equation}\label{dec_postPro}
p(\qx|\qy,\qH) \propto f_0(\qx)\prod\limits_{c\in\calC}f_c(\qx),
\end{equation}
with ${f_0}(\qx)=p(\qx)$ and ${f_c}(\qx) = \exp{\left(-\|\qy_c-\qH_c\qx\|^{2}/\sigma^{2}\right)}$ for $c\in\calC$.
In this manner, we represent \eqref{postPro} as a non-loopy factor graph with vector-valued nodes in Fig. \ref{F2A}.
This factor graph contains $C+1$ factor nodes $\{f_0,f_1,\ldots,f_C\}$ , and one variable node $\qx$.
Then, we restrict messages updated and transfered between nodes on the factor graph as Gaussian distributions.
In principle, Gaussian distributions can be fully characterized by their expectations and variances.
Thus, we only need to compute and propagate the expectations and variances of these messages.
On this basis, this process is named as expectation propagation.
Then, we approximate the a posteriori distribution \eqref{postPro} by computing and passing messages among the nodes in the factor graph (Fig. \ref{F2A}) in an iterative manner based on certain rules, and we obtain Algorithm \ref{A1}.
We elaborate the derivation of Algorithm \ref{A1} in Appendix A, while provide in the next subsection some operational explanations for the proposed detector.
\begin{figure}[!t]
	\centering
	\includegraphics[scale=0.4]{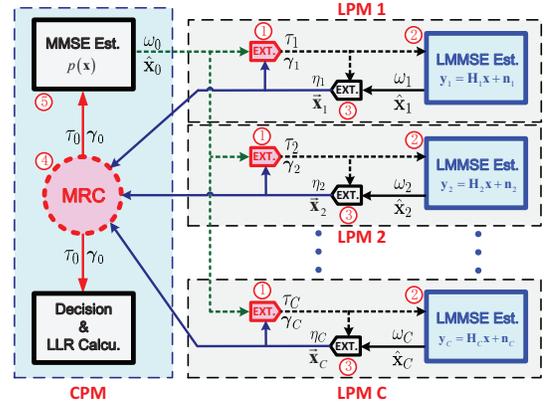}
	\caption{Block diagram of the proposed detector. The ``ext." block represents the extrinsic information computation.}\label{F12}
\end{figure}

\begin{Remark}
	Various algorithms proposed previously can be derived utilizing the EP principle for certain forms of factor graph, in the same manner as the framework presented in Appendix A.
	For example, the procedure in Appendix A can be followed to derive the original centralized EP MIMO detector \cite{Jespedes-2014TCOM} with the factor graph shown in Fig. \ref{F2B} by alternating between the selection of the factors $p(\qy|\qx,\qH)$ and $p(\qx)$.
	In other words, when $C=1$, Algorithm \ref{A1} presented in the next subsection is reduced to the centralized EP MIMO detector \cite{Jespedes-2014TCOM}.
	Algorithms proposed in \cite{Yuan-2014TWC,Rangan-2016Arxiv,JMA-2016SPL,JMa-2016Access} can be derived on the basis of the EP principle for a factor graph similar to Fig. \ref{F2B} as well.
	By introducing an auxiliary variable and a factor node corresponding to the linear transformation, the EP principle can be exploited to develop effective algorithms involving nonlinear measurements, such as \cite{CKWen-2016ISIT,Wang-2017JSAC,He-2018JSTSP,JMo-2018TSP}.
	Therefore, the EP principle will continue to help in developing effective algorithms for a great number of other applications, if we can describe the particular application scenario properly with a factor graph allowing for feasible computation of messages.
\end{Remark}

\begin{algorithm}[!t]
	\caption{EP-based Detector for subarray-based architecture \label{A1}}
	\small
	\KwIn{$\qy_c$ and $\qH_c$ for $c\in\calC$\;}
	\KwData{$\rqx_c = \qzero$ and $ \eta_c = 0$ for $c\in\calC$,  $\omega_{0}=E^{-1}_x$ and $\hat{\qx}_0=\qzero$\;}
	\Repeat{$\mathrm{a~certain~termination~criteria~holds}$}{
		\textbf{Parallel processing performed by each subarray ($c\in\calC$)}\\
		(1) The a priori mean and variance of $\qx$ for subarray $c$
		\begin{subequations}
			\begin{align}
			&\tau_{c} = \omega_{0}-\eta_c\label{pri_var_dec}\\
			&\qgamma_{c} = \tau^{-1}_{c}\left(\omega_{0}\hat{\qx}_0-\eta_c\rqx_c\right)\label{pri_m_dec}
			\end{align}
		\end{subequations}
		
		(2) The a posteriori mean and variance of $\qx$ for subarray $c$
		\begin{subequations}
			\begin{align}
			&\qSigma_c = {\left(\sigma^{-2} \qH_c^H\qH_c+ \tau_{c}\qI_K\right)}^{-1} \label{lmmse_var_dec}\\
			&\hat{\qx}_c = \boldsymbol{\Sigma}_c{\left(\sigma^{-2} \qH_c^H\mathbf{y}_c + \tau_{c}\qgamma_{c}\right)}\label{lmmse_mean_dec}\\
			&\omega_c = \left(K^{-1}\tr\left(\qSigma_c\right)\right)^{-1} \label{ave_mmse_dec}
			\end{align}
		\end{subequations}
		
		(3) The extrinsic mean and variance of $\qx$ for subarray $c$
		\begin{subequations} \label{eA1_a0304}
			\begin{align}
			&\eta_c = \omega_{c} - \tau_{c}\label{ext_var_dec}\\
			&\rqx_c = \eta^{-1}_c\left(\omega_{c}\hat{\qx}_c-\tau_{c}\qgamma_{c}\right) \label{ext_m_dec}
			\end{align}
		\end{subequations}
		
		\textbf{Combination and processing performed at CPM}\\
		(4) MRC combination:
		\begin{subequations}
			\begin{align}
			&\tau_{0} = \sum_{c=1}^C \eta_c\label{input_snr}\\
			&\qgamma_{0} = \tau^{-1}_{0}\sum_{c=1}^C\eta_c\rqx_c\label{input_obs}
			\end{align}
		\end{subequations}
		
		(5) MMSE estimation of $\qx$:
		\begin{subequations} \label{eA1_b0102}
			\begin{align}
			&\hat{x}_{0,k} = \Ex {\left\{ x_k | \gamma_{0,k},  \tau_{0} \right\}},~\text{for}~k\in\calK \label{post_mean_fus}\\
			&v_{0,k}= {\sf Var} {\left\{ x_k | \gamma_{0,k},  \tau_{0} \right\}},~\text{for}~k\in\calK \label{post_var_fus}\\
			&\omega_0 = \left(K^{-1}\sum_{k=1}^{K}v_{0,k}\right)^{-1}\label{ave_mmse_fus}
			\end{align}
		\end{subequations}	
	}
	\KwOut{$\tau_{0}$ and $\qgamma_{0}$}
\end{algorithm}

\subsection{Proposed Detector}

The block diagram of the proposed EP-based detector is illustrated in Fig. \ref{F12}, which provides detailed illustration for the block ``Local Det." in Fig. \ref{F1}.
The algorithm alternates between parallel computation in each LPM (Steps (1)$-$(3)), and combination and estimation in the CPM (Steps (4) and (5)).
Each LPM in parallel produces coarse estimates of $\qx$ based on their associated received vector, channel matrix, and feedback from the CPM of the last iteration.
Then, the extrinsic information, which is passed forward to the CPM (shown by the blue solid arrow in Figs. \ref{F1} and \ref{F12}), is calculated.
Subsequently, the CPM refines the estimate of $\qx$ considering the a priori distribution $p(\qx)$, which is broadcasted back to each LPM (shown by the green dotted arrow in Figs. \ref{F1} and \ref{F12}) for the next iteration. 
The entire procedure is summarized mathematically in Algorithm \ref{A1} with explicit expressions for Steps (1)$-$(5).
Then, we explain in detail the operations of the proposed algorithm.

We start by explaining the operations performed in the LPMs of each subarray.
At the beginning of the iterative process, $\tau_{c}$ and $\qgamma_{c}$ are initialized to be  $\qzero$ and $E^{-1}_x$ respectively.
At this moment, no extra information about $\qx$ is available except the mean and covariance of $\qx$.
Then, we compute the LMMSE estimation of $\qx$ with respect to (w.r.t.) the linear model in \eqref{dec_chan_model_ST} with the a priori mean and covariance matrix given by $\qgamma_{c}$ and $\tau^{-1}_{c}\qI_K$ respectively.
Following \cite[Theorem 12.1]{Kay1993} with some manipulations on the basis of the matrix inversion lemma, we can obtain the a posteriori covariance matrix and mean of $\qx$ by \eqref{lmmse_var_dec} and \eqref{lmmse_mean_dec}, respectively.
Next, we calculate the average a posteriori variance and take its reciprocal as $\omega_c$ in \eqref{ave_mmse_dec}.
Subsequently, we calculate the extrinsic information, which can be obtained by excluding the
a priori distribution of $\qx$ from its a posteriori distribution.
Following the Gaussian message combining rule \cite[(54) and (55)]{Loeliger-2007ProIEEE}, the extrinsic variance and mean can be given by \eqref{ext_var_dec} and \eqref{ext_m_dec} respectively. They are then transferred to the CPM via the feedforward link.

Subsequently, we turn to the operations performed by the CPM.
After collecting all extrinsic means and variances from each subarray, the CPM performs the maximum-ratio combining (MRC) expressed by \eqref{input_snr} and \eqref{input_obs}\footnote{\label{FN2}{ The name ``MRC" originates from the fact that Eqs. \eqref{input_snr} and \eqref{input_obs} are identical to the linear combination of $\{\rqx_c\}_{c=1}^{C}$ with minimum post-combination signal-to-noise ratio (SNR) by viewing $\rqx_c$ as AWGN observation of $\qx$ with noise power $\eta^{-1}_c$.}}.
The above MRC forms $\qgamma_{0}$ as an AWGN observation of $\qx$ with noise power $\tau^{-1}_{0}$, that is
\begin{equation}\label{AWGN_obs}
	\qgamma_{0}=\qx+\qn_0,
\end{equation}
where $\qn_0\sim\mathcal{CN}(\qzero,\tau^{-1}_{0}\qI)$.
Together with the knowledge that each entry of $\qx$ are equi-probably taken from the constellation $\calX$, we prepare to calculate the MMSE estimation of $\qx$ from \eqref{AWGN_obs}.
Equivalently, we compute the a posteriori mean and variance of $x_k$ for $k\in\calK$ w.r.t. the posterior probability
\[\mathrm{P}(x_k | \gamma_{0,k},\tau_{0})=\frac{\mathcal{CN}(\gamma_{0,k};x_k,\tau^{-1}_{0})\mathrm{P}(x_k)}{ \sum\limits_{ x\in\calX } \mathcal{CN}(\gamma_{0,k};x_k,\tau^{-1}_{0})\mathrm{P}(x) },\]
then the explicit expressions of $\hat{x}_{0,k}$ and $v_{0,k}$ can be given by
\begin{equation}\label{mmse_est_fus}
	\begin{aligned}
		\hat{x}_{0,k} &= \frac{{\sum\limits_{ x\in\calX } {x\, \mathcal{CN}\left(x;\gamma_{0,k},\tau^{-1}_{0}\right)} }}{{ \sum\limits_{ x\in\calX } { \mathcal{CN}\left(x;\gamma_{0,k},\tau^{-1}_{0}\right)} }},\\
		v_{0,k} &= \frac{{\sum\limits_{ x\in\calX } { {{\left| {x} \right|}^2}\mathcal{CN}\left(x;\gamma_{0,k},\tau^{-1}_{0}\right)} }}{{ \sum\limits_{ x\in\calX } { \mathcal{CN}\left(x;\gamma_{0,k},\tau^{-1}_{0}\right)} }} - {\left| \hat{x}_{0,k} \right|^2},
	\end{aligned}
\end{equation}
respectively. Finally, the estimate $\hat{\qx}_0$ and its corresponding average MSE computed in \eqref{ave_mmse_fus} are sent back to the LPMs of each subarray through the feedback link.
Next, each subarray calculates their a priori mean and variance of $\qx$ as \eqref{pri_m_dec} and \eqref{pri_var_dec} respectively, thereby starting the next iteration.
The above process is performed iteratively until a certain iteration stopping criterion is met.
The most convenient stopping criterion is the maximum iteration numbers.
After the convergence of the iteration, we obtain the approximated marginal posterior probability $\mathrm{P}(x_k \mid \qy,\qH)\approx\mathcal{CN}(x_k;\hat{x}_{0,k},v_{0,k})$.

When computing $\rqx_c$ in \eqref{ext_m_dec}, $\eta_c$ is divided.
The subsequent operations in \eqref{input_obs} and \eqref{pri_m_dec} include $\eta_c\rqx_c$.
To avoid redundant division and multiplication of $\eta_c$, in practical applications, we can compute $\rqx_c$ by
$\rqx_c = \omega_{c}\hat{\qx}_c-\tau_{c}\qgamma_{c}$.
Accordingly, the computation of $\qgamma_{0}$ and $\qgamma_{c}$ in \eqref{ext_m_dec} and \eqref{pri_m_dec} can be simplified to be
\[\qgamma_{0} = \frac{1}{\tau_{0}}\sum_{c=1}^C\rqx_c,~\qgamma_{c} = \frac{1}{\tau_{c}}\left(\omega_{0}\hat{\qx}_0-\rqx_c\right).\]

\subsection{Intuitive Verification of Convergence}
Firstly, we characterize the performance for the proposed EP-based detector based on the evolution technique.
Its basic idea is to select a few key parameters, which can be applied to characterize the statistical behavior of the iterative process and then to calculate the so-called transfer functions to track their evolution via iterations.
We can observe from Algorithm \ref{A1} that the final estimate of $\qx$ is produced from their AWGN observation $\qgamma_{0}$ and that its performance can be determined by the average noise power $\tau^{-1}_{0}$ from well-established formulas.
Therefore, the first parameter is selected as $\rho=\tau^{-1}_{0}$.
In addition, from \eqref{ext_var_dec} and \eqref{input_snr}, we find that $\tau_{0}$ and $\{\tau_{c}\}_{c=1}^{C}$ are mutually related in a recursive manner.
Hence they should also be selected as the parameters to be examined.
To enable concise expression, we denote $\nu_c=\tau^{-1}_{c}$ for $c\in\calC$.
Subsequently, we calculate transfer functions $\rho$ and $\nu_c$'s to show their evolution with the iteration, thereby inducing the following proposition.
\begin{Lemma}
	The evolution of parameters $\rho$ and $\nu_c$'s can be characterized recursively by
	\begin{subequations} \label{eq:SE_parameters}
		\begin{align}
		&\rho^{t}=\left(\sum_{c=1}^C\left(\frac{1}{\mse_c\left(\nu^t_c\right)}-\frac{1}{\nu^t_c}\right)\right)^{-1},\displaybreak[0]\label{eq:SE1}\\
		&\nu^{t+1}_c=\left(\frac{1}{\mse_0\left(\rho^{t}\right)}-\left(\frac{1}{\mse_c\left(\nu^t_c\right)}-\frac{1}{\nu^t_c}\right)\right)^{-1}~~\mathrm{for}~~c\in\calC,\label{eq:SE2}
		\end{align}		
	\end{subequations}
    which is initiated by $\nu^{1}_c=E_{x}$ for $\forall c\in\calC$, where the subscript $t$ denotes the iteration index, $\lambda_{i,c}$ is the $i$th eigenvalue of the matrix $\qH^{H}_c\qH_c$,
	$\mse_c\left(\nu^t_c\right)$ denotes the average MSE of the estimator \eqref{ave_mmse_dec} given $\qH_c$ and $\nu^t_c$	for each subarray $c\in\calC$, which is given by
	\begin{equation}\label{eq:MSE1}
	\mse_c\left(\nu^t_c\right)=\frac{1}{K}\sum\limits_{i=1}^{K}\frac{\sigma^2\nu^t_c}{\lambda_{i,c}\nu^t_c+\sigma^2},
	\end{equation}
	and $\mse_0\left(\rho^{t}\right)$ denotes the average MSE of the MMSE estimation of $x$ given its AWGN observation $r$ with average noise power $\rho^{t}$, which is given by
	\begin{equation}\label{eq:MSE2}
	 \mse_0\left(\rho^{t}\right)=\Ex_r\left\{\Ex_{x|r}\left[\left|x-\Ex_{x|r}\left[x|r\right]\right|^2\right]\right\}.
	\end{equation}
\end{Lemma}

\begin{IEEEproof}
	The transfer function of $\rho$ from $\nu_c$'s in \eqref{eq:SE1} can be calculated straightforwardly by substituting \eqref{ext_var_dec} into \eqref{input_snr}.
	Note that here we compute the average a posteriori variance $\omega_c$ of the estimator \eqref{ave_mmse_dec} by applying the singular-value decomposition (SVD) of $\qH_c$.
	The transfer functions of $\nu_c$'s from $\rho$ can be calculated similarly by substituting \eqref{ext_var_dec} into \eqref{pri_var_dec}.
	For a sufficiently large $N$, the average a posteriori variance $\omega_0$ in \eqref{ave_mmse_fus} can be computed compactly by \eqref{eq:MSE2} from the central limit theorem.
\end{IEEEproof}

The above evolution analysis has many interesting applications, such as intuitive convergence verification.
Proving the overall convergence of the proposed EP-based detector is generally difficult \cite{Minka-2005}.
However, the result of the evolution analysis provides us with the possibility to justify the convergence intuitively if we can show that the sequences $\{\rho^{t}\}$ and $\{\nu^{t}_c\}_{c=1}^{C}$ w.r.t. $t$ are bounded and monotonous, as shown in the proposition below.
Later in Section V, we will show through simulation results that the evolution analysis can provide precise performance prediction, and that the proposed algorithm converges.
\begin{Proposition}\label{evo_converg}
	The sequences $\{\rho^{t}\}$ and $\{\nu^{t}_c\}_{c=1}^{C}$ w.r.t. $t$ are bounded and monotonically decreasing.
\end{Proposition}

\begin{IEEEproof}
	See Appendix B.
\end{IEEEproof}

\subsection{Fixed Point Characterization}
In this subsection, we characterize the fixed points of the proposed EP-based detector.
In particular, we show that the fixed points are stationary points of a relaxed version of the KL divergence minimization distribution approximation problem.
We then clarify the relation of the fixed points to the replica prediction of the asymptotic MMSE performance.
Before proceeding, a simple consistency result, which is needed by subsequent discussions, is given by Lemma \ref{Fixed_Point}.
\begin{Lemma}\label{Fixed_Point}
	For any fixed point of Algorithm \ref{A1} with $\tau_{0}+\sum_{c=1}^{C}\tau_{c}>0$, we have
	\begin{subequations}%
		\begin{align}
		&\omega_{0}=\omega_{1}=\cdots=\omega_{C}=\omega:=\frac{1}{C}\left(\tau_{0}+\sum_{c=1}^{C}\tau_{c}\right),\displaybreak[0]\label{fix_point_var}\\
		&\hat{\qx}_0=\hat{\qx}_1=\cdots=\hat{\qx}_C=\hat{\qx}:=\frac{1}{C}\cdot\frac{\tau_{0}\qgamma_{0}+\sum_{c=1}^{C}\tau_{c}\qgamma_{c}}{\tau_{0}+\sum_{c=1}^{C}\tau_{c}}.\label{fix_point_m}
		\end{align}	
	\end{subequations}
\end{Lemma}

\begin{IEEEproof}
	Substituting \eqref{ext_var_dec} into \eqref{pri_var_dec}, we have $\omega_{0}=\omega_{c}$ for $c\in\calC$.
	Similarly, substituting \eqref{ext_m_dec} into \eqref{pri_m_dec}, and together with the fact that $\omega_{0}=\omega_{c}$, we have $\hat{\qx}_0=\hat{\qx}_c$ for $c\in\calC$.
	Then, \eqref{fix_point_var} and \eqref{fix_point_m} can be proved by combining \eqref{ext_var_dec} with \eqref{input_snr}, and combining \eqref{ext_m_dec} with \eqref{input_obs} respectively.
\end{IEEEproof}

The typical task of approximating a complicated distribution, such as \eqref{postPro}, is to minimize the Kullback-Leibler (KL) divergence between the approximated distribution and the original one over a given distribution family.
However, this minimization is generally intractable as it involves a search over a family of $K$-dimension distributions.
However, exploiting certain relaxations of the aforementioned KL divergence minimization problem allows for feasible solutions.
The following proposition reveals that the fixed points of the proposed detector characterizes the stationary point of a particular relaxed version of the KL divergence minimization problem.

\begin{Proposition}\label{free_ener_min}
	Denote the probability density functions $b_0(\qx)$, $\{b_c(\qx)\}_{c=1}^{C}$, and $q(\qx)$ parameterized by the common values $\omega$ and $\hat{\qx}$ of the fixed points shown in \eqref{fix_point_var} and \eqref{fix_point_m} as
	\begin{subequations}\label{stat_points}
		\begin{align}
			&b_0(\qx):=\frac{1}{Z_0\left(\qgamma_{0}\right)}p(\qx)\exp\left[-\tau_0\|\qx-\qgamma_{0}\|^2\right],\displaybreak[0]\label{Unbiased_Est_x}\\
			&b_c(\qx):=\frac{1}{Z_c\left(\qgamma_{c}\right)}\exp\left[-\frac{1}{\sigma^{2}}\|\qy_c-\qH_c\qx\|^{2}-\tau_c\|\qx-\qgamma_{c}\|^2\right],\displaybreak[0]\\
			&q(\qx):=\frac{1}{Z_q\left(\hat{\qx}\right)}\exp\left[-\omega\|\qx-\hat{\qx}\|^2\right],
		\end{align}
	\end{subequations}
	respectively, where $Z_0\left(\qgamma_{0}\right)$, $\{Z_c\left(\qgamma_{c}\right)\}_{c=1}^{C}$, and $Z_q\left(\hat{\qx}\right)$ are the normalization factor for their corresponding density functions.
	Then, $b_0(\qx)$, $\{b_c(\qx)\}_{c=1}^{C}$, and $q(\qx)$ are the stationary points of the optimization of the Bethe free energy subject to the moment-matching constraints, which can be expressed as
	\begin{subequations}\label{BFE_Min}
		\begin{align}
			&\min\limits_{b_0,b_1,\ldots,b_C}\max\limits_{q}~J\left(b_0,b_1,\ldots,b_C,q\right)\notag\\
			&~~~~~~~~~~~:=\sum\limits_{c=0}^{C}\KL\left(b_c\|e^{\log f_c(\qx)}\right)+CH(q)\label{Bethe_Free_Energy}\\
			&~~~~~\text{s.t.}~~~~\Ex_{b_c}(\qx)=\Ex_{q}(\qx),\displaybreak[0]\label{BFE_Min_Const1}\\
			&~~~~~~~~~~~~\frac{1}{K}\sum\limits_{k\in\calK}\Ex_{b_c}(|x_k|^2)= \frac{1}{K}\sum\limits_{k\in\calK}\Ex_{q}(|x_k|^2),\label{BFE_Min_Const2}
		\end{align}
	\end{subequations}
\end{Proposition}
where equality constraints \eqref{BFE_Min_Const1} and \eqref{BFE_Min_Const2} are set for $c=0,1,\ldots,C$, $\KL\left( { \cdot \left\|  \cdot  \right.} \right)$ denotes the KL divergence of two distributions and $H(\cdot)$ denotes the differential entropy.
In addition, $\omega$ and $\hat{\qx}$ satisfy that
\begin{subequations}\label{Moment_Matching}
	\begin{align}
	&\hat{\qx}=\Ex_{b_0}(\qx)=\Ex_{b_1}(\qx)=\cdots=\Ex_{b_C}(\qx)= \Ex_{q}(\qx),\\
	&\begin{aligned}
	\omega^{-1}&=\frac{1}{K}\sum\limits_{k\in\calK}\Ex_{b_0}(|x_k|^2)=\frac{1}{K}\sum\limits_{k\in\calK}\Ex_{b_1}(|x_k|^2)=\cdots\\
	&=\frac{1}{K}\sum\limits_{k\in\calK}\Ex_{b_C}(|x_k|^2)= \frac{1}{K}\sum\limits_{k\in\calK}\Ex_{q}(|x_k|^2).	
	\end{aligned}
	\end{align}
\end{subequations}
\begin{IEEEproof}
	See Appendix C.
\end{IEEEproof}

Now, we show how the KL divergence minimization problem is relaxed to be problem \eqref{BFE_Min}.
Under the constraint that $b_1(\qx)=\cdots=b_C(\qx)=q(\qx)$, the minimization of $J\left(b_1,\ldots,b_C,q\right)$ in \eqref{Bethe_Free_Energy} is equivalent to approximate $p(\qx|\qy,\qH)$ in \eqref{postPro} by a distribution with minimum KL divergence with $p(\qx|\qy,\qH)$, which is still intractable.
We further relax the constraint $b_1(\qx)=\cdots=b_C(\qx)=q(\qx)$ to be the moment-matching constraint in \eqref{BFE_Min_Const1} and \eqref{BFE_Min_Const2}, which requires a match in their first moments and average in their second moments.
Then, we provide an intuitive explanation of \eqref{AWGN_obs} as follows.

\begin{Remark}
	Consider the belief estimate $b_0(\qx)$ given by \eqref{Unbiased_Est_x}.
	If $\qgamma_{0}$ is modeled as a random vector, then \eqref{Unbiased_Est_x} implies that $b_0(\qx)$ represents the a posteriori of $\qx$ given $\qgamma_{0}$, namely, $p\left(\qx|\qgamma_{0}\right)$.
	Then, from the Bayes rule, we can express the likelihood function as
	\[p\left(\qgamma_{0}|\qx\right)=\frac{p\left(\qgamma_{0}\right)}{Z_0\left(\qgamma_{0}\right)}\exp\left[\tau_0\|\qx-\qgamma_{0}\|^2\right],\]
	where $p\left(\qgamma_{0}\right)$ denotes the marginal distribution $p\left(\qgamma_{0}\right)=\int p\left(\qx|\qgamma_{0}\right)p\left(\qx\right)\mathrm{d}\qx$.
	An admissible choice of $p\left(\qgamma_{0}\right)$ that satisfies $p\left(\qgamma_{0}\right)\geq0$, $\int p\left(\qgamma_{0}\right)\mathrm{d}\qgamma_{0}=1$, and $\int p\left(\qgamma_{0}|\qx\right)\mathrm{d}\qgamma_{0}=1$ is $p\left(\qgamma_{0}\right)=\pi^{-N}\tau^N_0Z_0\left(\qgamma_{0}\right)$.
	Then, we have $p\left(\qgamma_{0}|\qx\right)=\calCN\left(\qgamma_{0};\qx,\tau_0\qI\right)$.
	In this case, we can interpret $\qgamma_{0}$ as an unbiased estimate of $\qx$ Gaussian estimation error of variance $\tau_0$.
	On this basis, we can calculate the log-likelihood
	ratio (LLR) through $\qgamma_{0}$ and $\tau_0$ for soft decoding.
	In addition, the hard decision can also be performed by determining $x_k\in\calX$ for each $k\in\calK$ with the shortest distance to $\gamma_{0,k}$, as we do in Algorithm \ref{A1}.
\end{Remark}

Next, we characterize the asymptotic performance of the proposed EP-based detector as the following proposition.
\begin{Proposition}\label{fixed_point_equ}
	When $\qH$ is a random matrix with i.i.d. elements, for any fixed point $\omega$ and $\tau_{0}$ of proposed EP-based detector, it holds that
	\begin{equation}\label{Replica_Result}
	\tau_{0}=R_{\sigma^2\qH^H\qH}(-\omega^{-1}),~~\omega^{-1}=\frac{1}{K}\sum\limits_{k\in\calK}\Ex_{b_0}(|x_k|^2),
	\end{equation}
	under the large system regime where $N,N_1,N_2,\ldots,N_C,K\to\infty$ with $K/N$ being a fixed constant.
\end{Proposition}

\begin{IEEEproof}
	See Appendix D.
\end{IEEEproof}

\begin{Remark}
	We find that \eqref{Replica_Result} is identical to the fixed-point equation characterization
	of the asymptotic MSE performance for \eqref{postPro} derived from the replica method \cite[Eq. (17)]{Tulino-2013TIT}.
	Therefore, when the channel matrix $\qH$ contains i.i.d. entries, the proposed EP detector can potentially achieve the MMSE performance in certain asymptotic and random regimes.
	In other words, under any number and size of subarray, the proposed detector will produce similar performance.
	Furthermore, simulation results show that even when $\qH$ is non-i.i.d., reducing $N_c$ does not result in significant performance loss.
	As discussed in Section IV-B, including a small number of antennas in each subarray allows for simple-form computation of some matrix operations.
	The above observations inspire us to parallelize the complex operation into several simple-form computations, giving rise to the hierarchical implementation architecture proposed in Section IV-B.
\end{Remark}

\section{Efficient implementation-related issues}

In this section, we shift our discussion to the engineering perspective. 
In Subsection IV-A, we demonstrate that the computational complexity and the amount of data required to be transfered from each subarrays to the CPM can be reduced by using the trimmed local channel matrices.
Then, in Subsection IV-B, we propose a hierarchical architecture that enables the parallelism improvement.
In Subsection IV-C, we modify the proposed EP-based detector so that only one feedforward from subarrays to the CPM is required in the entire iteration procedure.
Finally in Subsection IV-D, we provide complexity comparison.

\subsection{Implementation of Algorithm \ref{A1} based on trimmed subarray model}

The principal advantage of implementing Algorithm \ref{A1} based on the trimmed subarray model \eqref{dec_chan_model_NST} is the reduction in computational complexity and the amount of data needed to transfer from each subarray to the CPM.
In this case, Step (2) in Algorithm \ref{A1} produces $\hat{\qx}_c$ as the LMMSE estimate of $\qx_c$, which is given by
\begin{equation}\label{lmmse_est_dec_NS}
	\begin{aligned}
		\qSigma_c &= {\left(\sigma^{-2} \tilde{\qH}_c^H\tilde{\qH}_c+ \tau_{c}\qI_{K_c}\right)}^{-1},\\
		\hat{\qx}_c &= \boldsymbol{\Sigma}_c{\left(\sigma^{-2} \tilde{\qH}_c^H\mathbf{y}_c + \tau_{c}\qgamma_{c}\right)},\\
		\omega_c &= \left(K^{-1}_c\tr\left(\qSigma_c\right)\right)^{-1}.
	\end{aligned}
\end{equation}
Then, the number of complex-value multiplication required in producing $\hat{\qx}_c$ is reduced from $K(N_c+K)$ to $K_c(N_c+K_c)$,
and the number of complex-value multiplication required in computing $\qSigma_c$ using the method presented in Algorithm \ref{A2} is reduced from $N_cK(2K+1)$ to $N_cK_c(2K_c+1)$.
Moreover, the number of complex numbers required to transfer to the CPM, namely the dimension of $\rqx_c$, is reduced from $K$ to $K_c$.  
Typically, $K_c$ is much smaller than $K$.
Therefore, the computational complexity and information exchange overhead can be considerably reduced.
Accordingly, the MRC combination performed in Step (4) of Algorithm \ref{A1} needs to be modified as 
\[\tau_{0,k} = \sum_{c\in\calC_k} \eta_c,~~
\gamma_{0,k} = \frac{1}{\tau_{0,k}}\sum_{c\in\calC_k}\eta_c\rightvec{x}_{c,k}\]
for $k\in\calK$, where $\calC_k$ denotes the subset of $\calC$ satisfying the condition that if $c\in\calC_k$, then $k\in\calK_c$.
In the computations of $\hat{x}_{0,k}$ and $v_{0,k}$ using \eqref{mmse_est_fus}, we substitute $\tau_{0}$ by $\tau_{0,k}$.
In addition, we compute $\qgamma_{c}$ as  $\qgamma_{c}=\tau^{-1}_{c}\left(\omega_{0}[\hat{\qx}_0]_{\calK_c}-\eta_c\rqx_c\right)$.
Except for the steps mentioned above, the expressions for the other steps are unchanged.

\subsection{Hierarchical Implementation Architecture}
The primary bottleneck in the practical implementation of the proposed EP-based detector is that the matrix inversion in \eqref{lmmse_var_dec}, whose complexity is in the order of $K^3$ considering full subarray model, or $K_{c}^3$ considering trimmed subarray model, is required in every iteration.
Under a crowded scenario, this complexity is unaffordable.
However, when the antenna number in each subarray is small, a simple computation of the matrix inversion can be performed.
Consider the extreme case, $N_c=1$, in which a computationally efficient formula is available for the matrix inversion.
In this case, $\qH_c$ is the $c$-th row of $\qH$.
The matrix  $\qH^H$ can be blocked as $\qH^H=[\bar{\qh}_1,\bar{\qh}_2,\ldots,\bar{\qh}_C]$.
Then $\qSigma_c$ in \eqref{lmmse_var_dec} can be calculated in a very simple form shown below,
\begin{equation}\label{simp_inv}
\qSigma_c=\tau^{-1}_{c}\qI-\frac{\tau^{-2}_{c}\sigma^{-2}\bar{\qh}_c\bar{\qh}^H_c}{1+\tau^{-1}_{c}\sigma^{-2}\bar{\qh}^H_c\bar{\qh}_c},
\end{equation}
which involves the multiplication of a column and a row vector divided by a scalar.
When $N_c$ is adequately small\footnote{What number of $N_c$ can be called adequately small depends on the computational speed of the component in which the algorithm is implemented.}, the matrix inversion can be computed by recursively using the above formula for $N_c$ times by exploiting the following observation:
\[\qSigma_c={\left(\tau_{c}\qI_K + \sigma^{-2}\sum\limits_{j=1}^{N_c}\bar{\qh}_{j,c}\bar{\qh}^H_{j,c}\right)}^{-1}.\]
We let $\qA_{c,0}=\tau_{c}\qI_K$ and $\qA_{c,j}=\tau_{c}\qI_K + \sigma^{-2}\sum_{i=1}^{j}\bar{\qh}_{i,c}\bar{\qh}^H_{i,c}$.
Evidently, $\qSigma_c=\qA^{-1}_{c,N_c}$ and $\qA_{c,j}=\qA_{c,j-1}+\sigma^{-2}\bar{\qh}_{j,c}\bar{\qh}^H_{j,c}$.
If $\qA^{-1}_{c,j-1}$ is given, $\qA^{-1}_{c,j}$ can be then obtained in the same manner as \eqref{simp_inv}.
On the basis of the above observation, $\qSigma_c$ can be calculated recursively using the procedure given by Algorithm \ref{A2}.

\begin{algorithm}[!th]
	\caption{Recursive computation of $\qSigma_c$ \label{A2}}
	\small
	\KwIn{$\tau_{c}$, $\sigma^{2}$ and $\qH_c$\;}
	\KwData{$\qA_{c,0}=\tau_{c}\qI_K$\;}
	\For{$j=1,2,\ldots,N_c$}{
		\[\qA^{-1}_{c,j}=\qA^{-1}_{c,j-1}-\frac{\sigma^{-2}\qA^{-1}_{c,j-1}\bar{\qh}_{c,j}\bar{\qh}^H_{c,j}\qA^{-1}_{c,j-1}}{1+\sigma^{-2}\bar{\qh}^H_{c,j}\qA^{-1}_{c,j-1}\bar{\qh}_{c,j}}\]	
	}
	\KwOut{$\qSigma_c=\qA^{-1}_{c,N_c}$}	
\end{algorithm}

If $N_c$ is a very small number, such as 1, 2, or 4, and $N$ is relatively large, such as 512, the number of subarrays is excessive.
In this condition, the amount of data that needs to be interchanged through the interface between subarrays and the CPM will be large.
In some cases, $N_c$ is fixed by hardware constraint, which cannot be adapted flexibly.
Then, the above strategy to make $N_c$ small is not feasible.
To keep $C$ a reasonable number but let the above strategy work, we propose to implement the proposed algorithm in a hierarchical manner by further dividing each subarray into several secondary subarrays.
For example, we divide $N_c$ antennas in the $c$-th subarray into $M_c$ secondary subarrays.
Then, we denote the antenna number in the $c'$-th secondary subarray by $N_{c,c'}$.
Although $N_c$ may be a relatively large number, we can set $N_{c,c'}$ to be small numbers such as 1, 2, or 4.
The LPM of each subarray assigns $M_c$ parallel computation units for each secondary subarray.
Then $\qy_c$ and $\qH_c$ can be further blocked as $\qy_c = [\qy_{c,1}^T, \qy_{c,2}^T, \ldots, \qy_{c,M_c}^T]^T$ and $\qH_c = [\qH_{c,1}^T,\qH_{c,2}^T,\ldots,\qH_{c,M_c}^T]^T$ respectively.
In each iteration, each secondary subarray performs Steps (1) - (3) of Algorithm \ref{A1}, substituting
the subscripts of all variables with subscript '$_c$' by '$_{c,c'}$'.
After the extrinsic information of each secondary subarray $\eta_{c,c'}$ and $\rqx_{c,c'}$ are obtained, each subarray combines them as its extrinsic information to deliver to the CPM as
\[\eta_{c} = \sum_{c'=1}^{M_c} \eta_{c,c'},~~\rqx_{c} =  \frac{1}{\eta_{c}}\sum_{c'=1}^{M_c}\eta_{c,c'}\rqx_{c,c'}.\]
After the feedback of $\omega_{0}$ and $\hat{\qx}_0$ from the CPM, each subarray broadcasts them to its secondary subarrays for the next iteration.
The above hierarchical operation is equivalent to keeping a small number $N_c$ of antennas in each subarray.
Notably, the above discussion is for the algorithm corresponding to the full subarray model and the same way can be directly applied for the algorithm for trimmed subarray model.

\begin{table*}[!t]
	\renewcommand{\arraystretch}{1.2}
	\centering
	\caption{Complexity Comparison\label{T3}}
	\begin{tabular}{c l c c c}\hline\hline
		\multirow{2}*{{Algorithm}} & ~ &  \multicolumn{2}{c}{{Each Parallel LPM}}  & \multirow{2}*{CPM} \\ \cline{3-4}
		~ & ~ &{Full $\qH_c$} & {Trimmed $\tilde{\qH}_c$} &  ~ \\
		\hline
		\multirow{4}*{Algorithm \ref{A1}} & \multirow{2}*{Mult.} & $8N_cTK(K+1)+$ & $8N_cTK_c(K_c+1)+$  & $CT(1+2K)+$\\ 
		~ & ~ & $6TK(K+2)$ & $6TK_c(K_c+2)$ & $K(T-1)(7|\calX|+2)$ \\ \cline{2-5}
		~ & Exp.& 0 & 0 & $K|\calX|(T-1)$\\
		~ & Trans.& $T(2K+1)$ & $T(2K_c+1)$ & $(2K+1)(T-1)$ \\
		\hline
		\multirow{4}*{One feedforward} &\multirow{2}*{Mult.} & $TK[8N_c(K+1)+2(4K+3)]$ & $TK_c[8N_c(K_c+1)+2(4K_c+3)]$ & \multirow{2}*{$C(1+2K)$} \\ 
		~ & ~ & $+K(T-1)(7|\calX|+6)$ & $+K_c(T-1)(7|\calX|+6)$ & ~\\ \cline{2-5}
		~ & Exp.& $KT|\calX|$ & $K_cT|\calX|$ & 0\\
		~ & Trans.& $2K+1$ & $2K_c+1$ & 0 \\
		\hline
		\multirow{2}*{Centralized EP} & Mult. & \multicolumn{3}{c}{$TK[8N(K+1)+2(4K+3)]+K(T-1)(7|\calX|+6)$} \\
		~ & Exp.& \multicolumn{3}{c}{$K|\calX|(T-1)$} \\
		\hline\hline
	\end{tabular}
\end{table*}

\subsection{Modified EP-based detector requiring one feedforward}
	
Another practical challenge is the need of multiple feedforward ($\{\eta_c\}_{c=1}^{C}$ and $\{\rqx_c\}_{c=1}^{C}$ from each subarray to the CPM) and feedback ($\omega_0$ and $\hat{\qx}_0$ from the CPM to each subarray) in every iteration for the information exchange between the subarrays and the CPM.
However, for application scenarios where the hardware entities of the subarrays and the CPM are placed distantly and connected by dedicated fronthaul links, such frequent iterative information exchange suffers from high interconnect latency. 
To address this challenge, we modify the proposed EP-based detector so that the entire iterative procedure requires only one feedforward.
To achieve this aim, two potential schemes are presented as follows.
\begin{itemize}
	\item For each received signal $\qy$, we perform only one iteration of Algorithm \ref{A1} to obtain the final output.
		
	\item Each subarray performs the entire procedure of Algorithm \ref{A1} (or the centralized EP algorithm \cite{Jespedes-2014TCOM}) for their local input-output relationship $\qy_c = \tilde{\qH}_c \qx_c + \qn_c$ in parallel.
	After convergence, each subarray transfers the estimate of $\qx$ and the corresponding variance
	(corresponding to the ``output" line in Algorithm \ref{A1}) to the CPM for MRC combination (in the same form of Eqs. \eqref{input_snr} and \eqref{input_obs}) as the final output.	
\end{itemize}
	
The first scheme is the most straightforward idea to achieve one feedforward.
However, it prevents performance improvement via iterations inherently provided by EP-based algorithms, which results in performance degradation.
This observation motivates us to retain the iteration and propose the second scheme.
The drawback of this scheme is that the information exchange among subarrays is disabled.  
Notably, when non-stationarity occurs, each subarray serves only a small portion of active users, and the overlap among sets of users served by different subarrays is considerably small.
Therefore, disabling information among subarrays may not result in a significant performance loss.

\subsection{Implementation Complexity}

Table \ref{T3} provides quantitative results in assessing implementation complexity of the proposed subarray-based detector.
We compare the complexity of Algorithm \ref{A1}, one feedforward architecture proposed in Section IV-C, and decentralized EP detector proposed in \cite{Jespedes-2014TCOM}. 
The complexity of the operations performed in LPM and CPM are counted separately.
Operations that implement in a single LPM are counted because all LPMs simultaneously perform parallel computations.
Table \ref{T3} lists the total number of real-valued multiplications, exponential operations, and real numbers exchanging between LPMs and the CPM (denoted by Mult., Exp. and Trans., respectively) required for performing $T$ iterations.
We also compare the complexity when full channel matrix $\{\qH_c\}_{c=1}^{C}$ and trimmed channel matrix $\{\tilde{\qH}_c\}_{c=1}^{C}$ are utilized by each LPM.
Table \ref{T3} shows that the required number of multiplications in each LPM is proportional to $K^2$ or $K^2_c$, and the dimension of data required to transfer from each LPM to the CPM is proportional to $K$ or $K_c$. 
Due to the non-stationarity, $K_c$ is typically considerably smaller than $K$.
Hence, the sparsity in local channel matrices reduces computational complexity and information exchange overhead.
In addition, one feedforward architecture significantly reduces the information exchange overhead compared with Algorithm \ref{A1}, thereby reducing the latency caused by frequent information exchange. 
Therefore, the one feedforward architecture is suitable for scenarios where the hardware entities corresponding to different subarrays and the CPM are distributed over a large area and interconnected by capacity limited fronthaul links.

\section{Simulation Results}
In this section, we demonstrate numerical results for performance evaluation and comparison.
We take the uncoded bit error rate (BER) as the performance metric using Monte Carlo simulations of 10,000 independent channel realizations.
Without the loss of generality, we set the number of antennas in each subarray $N_c$ to be identical in our simulations.
In addition, we mark the iteration numbers in the figures below to provide clues about the computational delays of each algorithm.

\subsection{Stationary Case}
In this subsection, we simulate the basic stationary case where a moderate number of antennas are compactly installed.
Various algorithms proposed for the full subarray model \eqref{dec_chan_model_ST} are evaluated and compared to examine their lossless performance and verify the validity of our analysis.
The stationary case corresponds to compact antenna deployment.
Hence, the correlation among antennas are considered.
Since stationary case corresponds to compact antenna deployment, we will consider correlation among antennas here.
We generate the channel matrix in our simulations as $\qH=\qSigma_R^{\frac{1}{2}}\qH_R$,
where $\qH_R$ is the Rayleigh fading matrix with i.i.d. Gaussian entries with zero mean and variance $K^{-1}$,
and $\qSigma_R$ characterizes the correlation among the elements of the antenna array, whose $(i,j)$-th entry is set to be $\kappa^{|i-j|}$, where $\kappa$ will be called correlation coefficients subsequently.
Moreover, we set the number of antennas $N=64$ and the number of users $K=16$ and elements of $\qx$ are drawn independently from the equiprobable 16-QAM constellation.

\begin{figure}[!t]
	\centering
	\subfloat[$N_c=1$]{\includegraphics[scale=0.6]{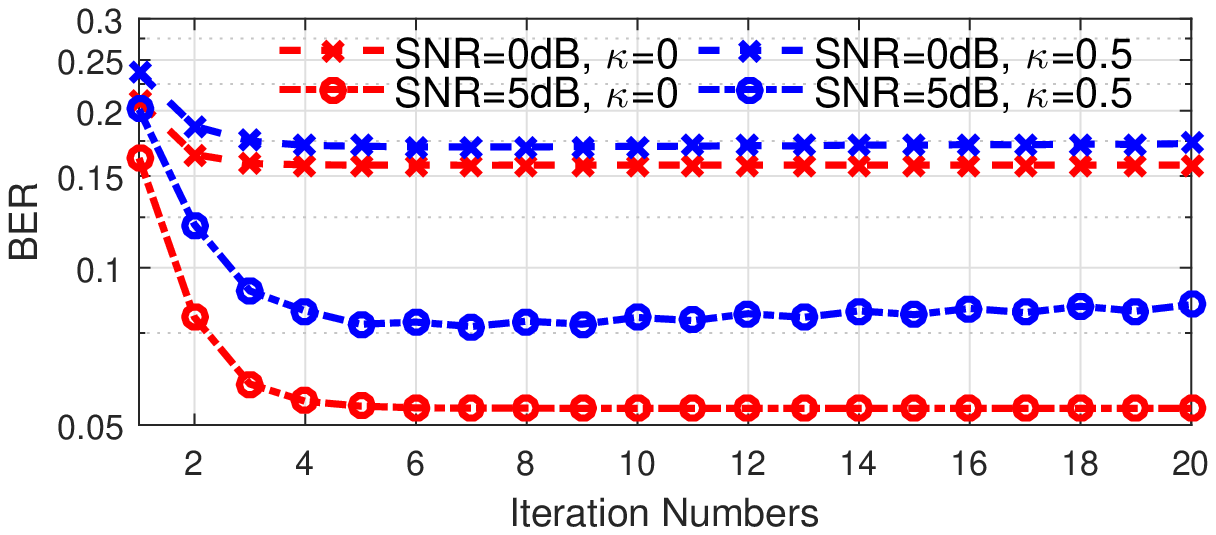}\label{F7A}}\\\vspace{-10pt}
	\subfloat[$N_c=2$]{\includegraphics[scale=0.6]{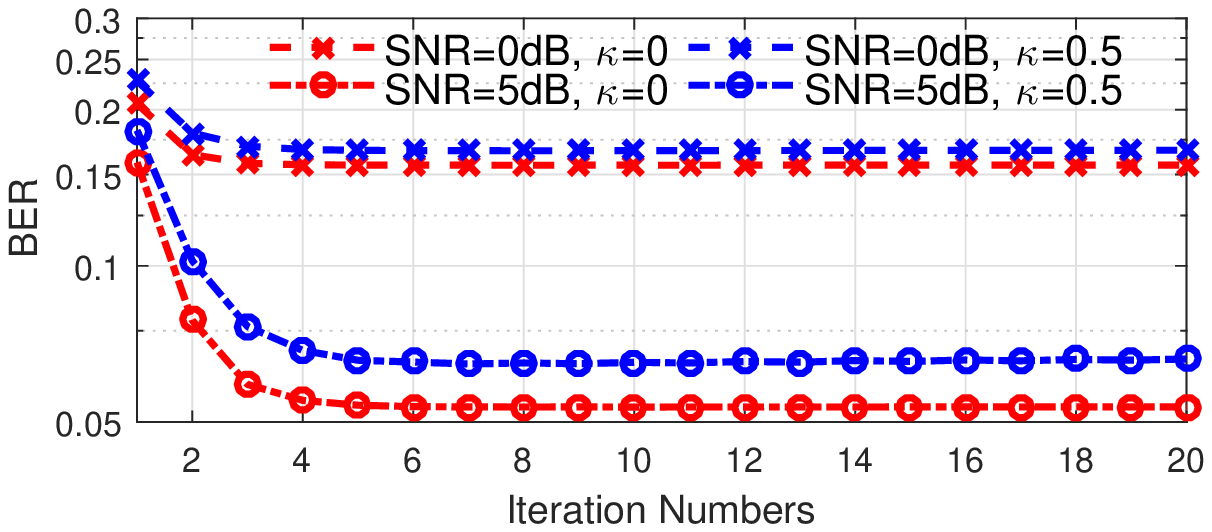}\label{F7B}}\\\vspace{-10pt}
	\subfloat[$N_c=4$]{\includegraphics[scale=0.6]{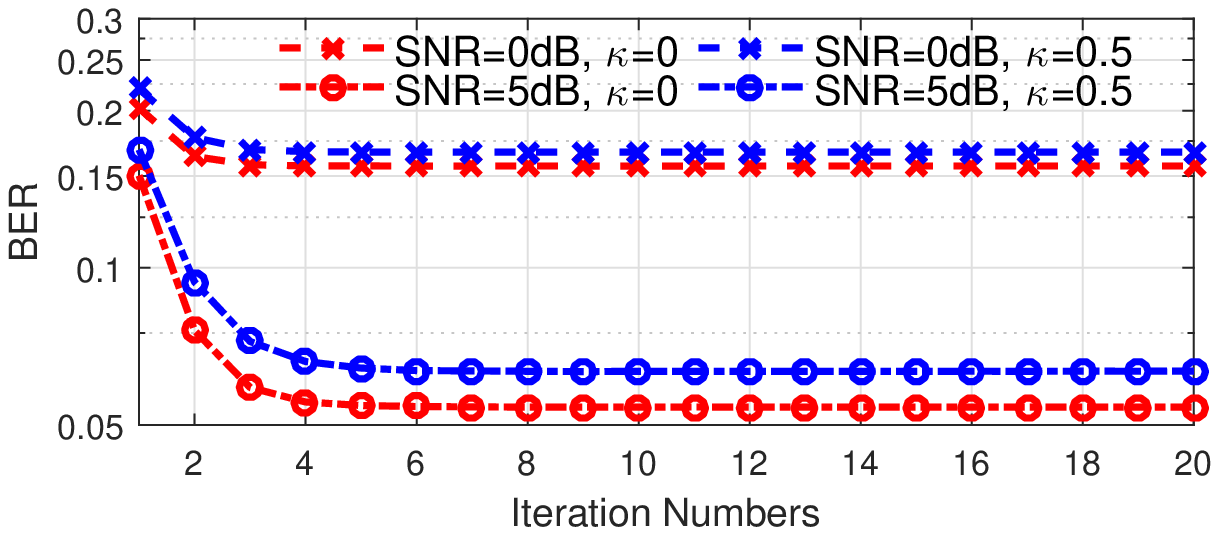}\label{F7C}}\\\vspace{-10pt}
	\subfloat[$N_c=16$]{\includegraphics[scale=0.6]{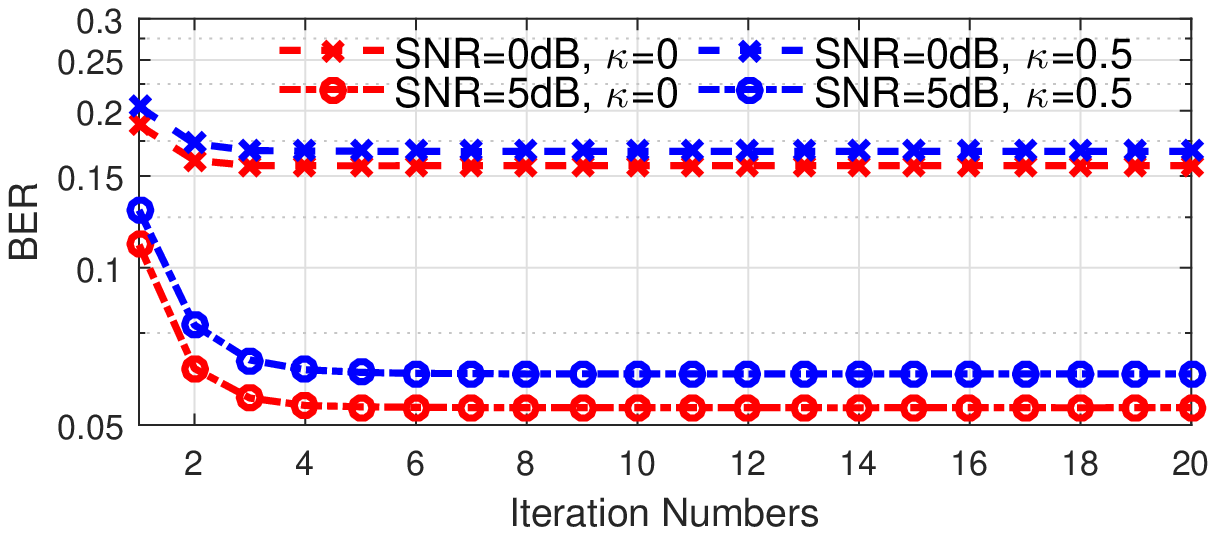}\label{F7D}}
	\caption{BERs versus the iteration numbers of the proposed
		EP detector (namely, Algorithm \ref{A1}) under different values of $N_c$ and $\kappa$ and SNR of 0 and 5 dB.\label{F7}}
\end{figure}

We first present the uncoded BER provided by Algorithm \ref{A1} versus the iteration numbers under different numbers of $N_c$ in Figs. \ref{F7A} - \ref{F7D}.
The results of Rayleigh channel ($\kappa=0$) and correlated channel ($\kappa=0.5$) are shown.
We observe from these figures that in all cases, the values of BER initially reduce and then remain stable with only minor fluctuations.
This observation alleviates our worries about the possible numerical instability of the proposed algorithm with a small number of antennas in each subarray ($N_c=$1,2 and 4) and the correlated channel.
The results in Fig. \ref{F7} justify our analysis in Section III-B.
From Fig. \ref{F7}, we also find that, with smaller number of antennas in each subarray, one or two more iterations are required to attain the numerical stability.
Specifically, when $N_c=$ 1, 2, or 4, the numerical stability is attained with 5 or 6 iterations, whereas with larger $N_c=$ 16, 4 iterations are enough to attain the numerical stability.
The decrease in BER is minor in the last few iterations.
Therefore, we recommend performing 3 or 4 iterations rather than waiting until stability is achieved completely in the practical applications.

\begin{figure}[!t]
	\centering
	\subfloat[Performance of 1 iteration]{\includegraphics[scale=0.6]{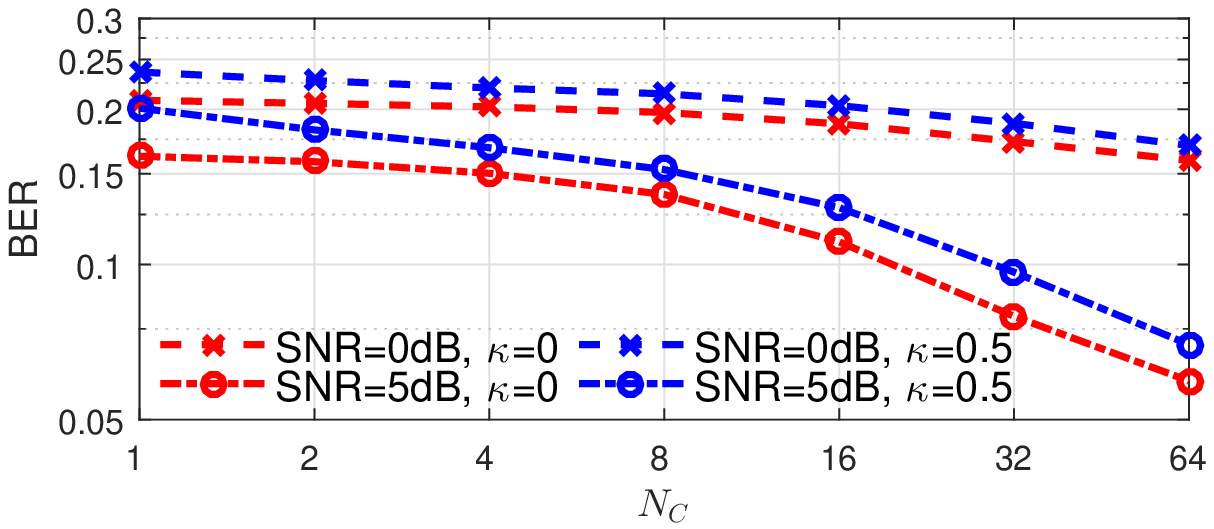}\label{F8A}}\\\vspace{-10pt}
	\subfloat[Performance after numerical stability]{\includegraphics[scale=0.6]{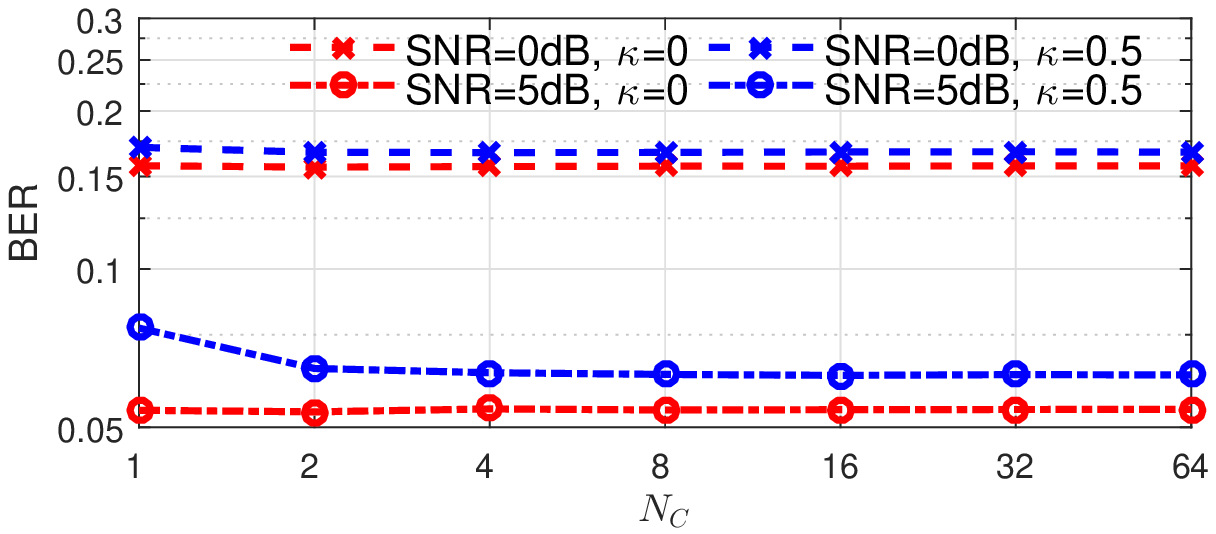}\label{F8B}}
	\caption{BERs versus the size of each subarray $N_c$ of the proposed
		EP detector (namely, Algorithm \ref{A1}) under $\kappa$ of 0 and 0.5, and SNR of 0 and 5 dB.\label{F8}}
\end{figure}

To examine the impacts of the value of $N_c$ on the performance of one iteration and that after stability, we provide BER versus $N_c$ with different iteration numbers in Fig. \ref{F8}, where we include the performance of seven iterations as that after stability.
As implied by Fig. \ref{F8A}, larger $N_c$ leads to better BER performance when one iteration is performed.
It finding is natural because one iteration of Algorithm \ref{A1} is equivalent to conducting MRC for the LMMSE detection for each subarray without any information exchange among different clusters, thereby improving the performance.
Then, in the application scenario where only one iteration is allowed, having more antennas in each subarray is favorable.
However, different results can be observed for the performance after stability.
For the Rayleigh channel, the partition into subarrays does not result in performance loss compared with centralized case, namely, $N_c=64$, thereby justifying the results revealed in Proposition \ref{fixed_point_equ}.
Meanwhile, in the correlated channel, small $N_c$ values give rise to slight performance loss compared with larger value of $N_c$.
We can conclude from the above observations that the computational complexity of matrix inversion can be alleviated by partitioning into subarrays with a small number of antennas in each of them at a low cost of performance degradation.

\begin{figure}[!t]
	\centering
	\subfloat[$\kappa=0$]{\includegraphics[scale=0.55]{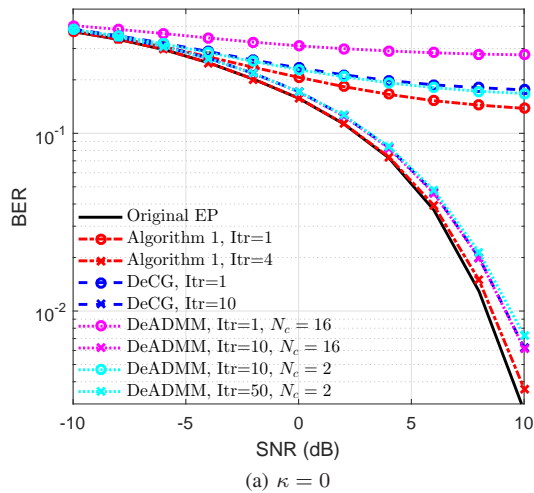}\label{F10A}}\\\vspace{-10pt}
	\subfloat[$\kappa=0.5$]{\includegraphics[scale=0.55]{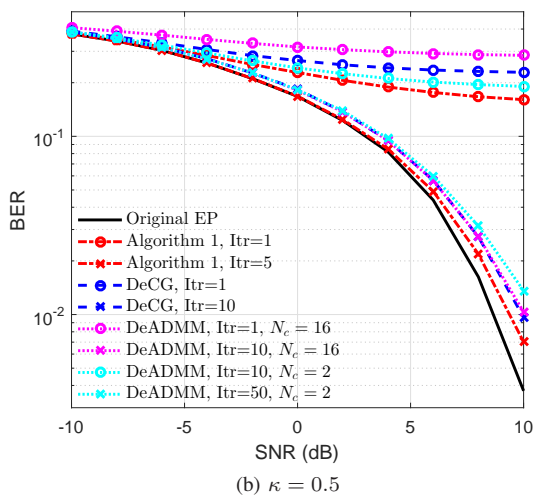}\label{F10B}}
	\caption{Comparison of the BER performances of various
		detectors proposed for the full subarray model under $N_c=2$, and $\kappa=0$ and 0.5.\label{F10}}
\end{figure}

Then, in Fig. \ref{F10}, we compare the performance of Algorithm \ref{A1} with that of various algorithms under full subarray model, including decentralized conjugate gradient (DeCG) and alternating direction method of multipliers (DeADMM) proposed in \cite{Studer-2017JESTCS}\footnote{In our simulation, we used part of the code downloaded from \url{https://github.com/VIP-Group/DBP}}, and the original centralized EP MIMO detector proposed in \cite{Jespedes-2014TCOM}.
To facilitate parallel simple computation of matrix inversion using Algorithm \ref{A2}, we set $N_c=2$, which yields a similar performance as that under larger $N_c$ as shown in Fig. \ref{F8B}.
In particular, the performances achieved with one iteration are emphasized because performing one iteration takes up the least computational resource and requires only one feedforward.
We find from Fig. \ref{F10} that with one iteration, Algorithm \ref{A1} performs better than DeGC and DeADMM.
Similarly, after attaining numerical stability, Algorithm \ref{A1} also outperforms DeGC and DeADMM.
Moreover, we observe that the performance of Algorithm \ref{A1} approaches that of the centralized EP MIMO detector.
This observation is consistent with the results shown in Fig. \ref{F8}.
Finally, when implementing DeADMM, we tend to choose larger $N_c$, as we observe that when $N_c=2$, DeADMM requires several times number of iterations to attain the same performance as that when $N_c=16$.
Therefore, choosing smaller $N_c$ gives rise to further computational delay.

\subsection{Non-stationary Case}

In this subsection, we consider a linear antenna array for our simulation setup, where the length of array is 250 meters with $N=512$ antennas, and $K=16$ active users have the same vertical distance to the array\footnote{\label{FN1}{Here, we follow the simulation setup in \cite{Amiri-2018GCW} to guarantee fair performance comparison with the algorithm proposed in \cite{Amiri-2018GCW}.}} and are uniformly distributed along the array.
Under this setting, the received energy from each user varies among different portions of the antenna array, and spatial non-stationarity occurs.
We then compare the performances of the algorithms that are proposed for the trimmed subarray model.
On this basis, we generate the channel matrix as $\qH=\qD\odot\qH_R$,
where $\qH_R$ is the Rayleigh fading matrix,
and $\qD$ contains the large-scale fading factors whose $(i,j)$-th entry is given by $d^{-1}_{ij}$ with $d_{ij}$ denoting the distance between $i$-th antenna and $j$-th user.
The process of channel matrix acquisition is not simulated, therefore, we utilize \cite[Algorithm 2]{Amiri-2018GCW} for the full channel matrix $\qH$ with the power threshold set to 0.9 to determine which part of users is served by each subarray.
In this manner, we generate trimmed channel matrices $\{\tilde{\qH}_c\}$ used in our simulations.

\begin{figure}[!t]
	\centering
	\subfloat[SNR=-5 dB]{\includegraphics[scale=0.6]{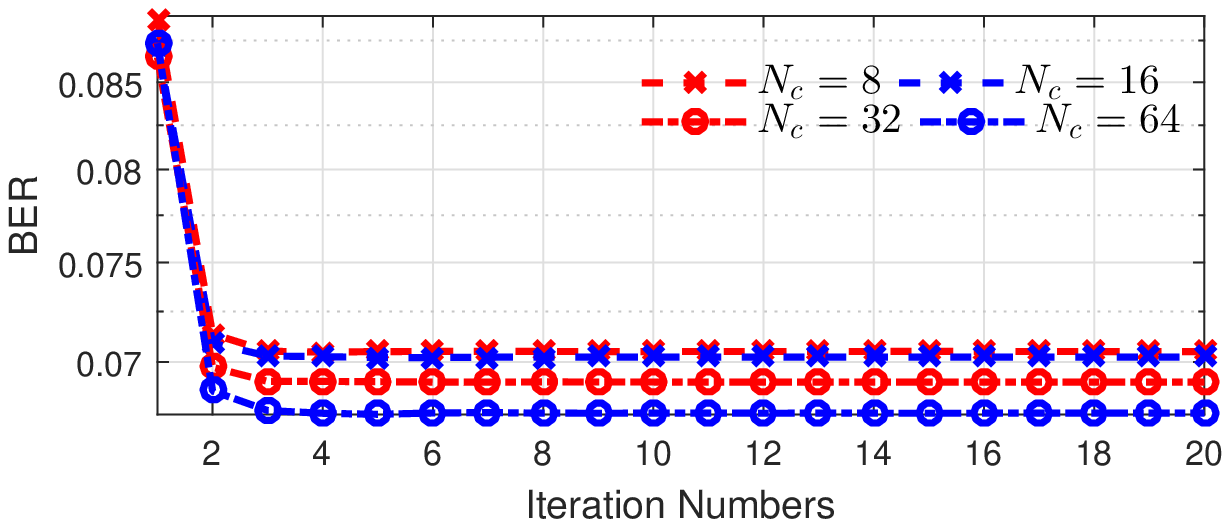}\label{F9A}}\\\vspace{-10pt}
	\subfloat[SNR=0 dB]{\includegraphics[scale=0.6]{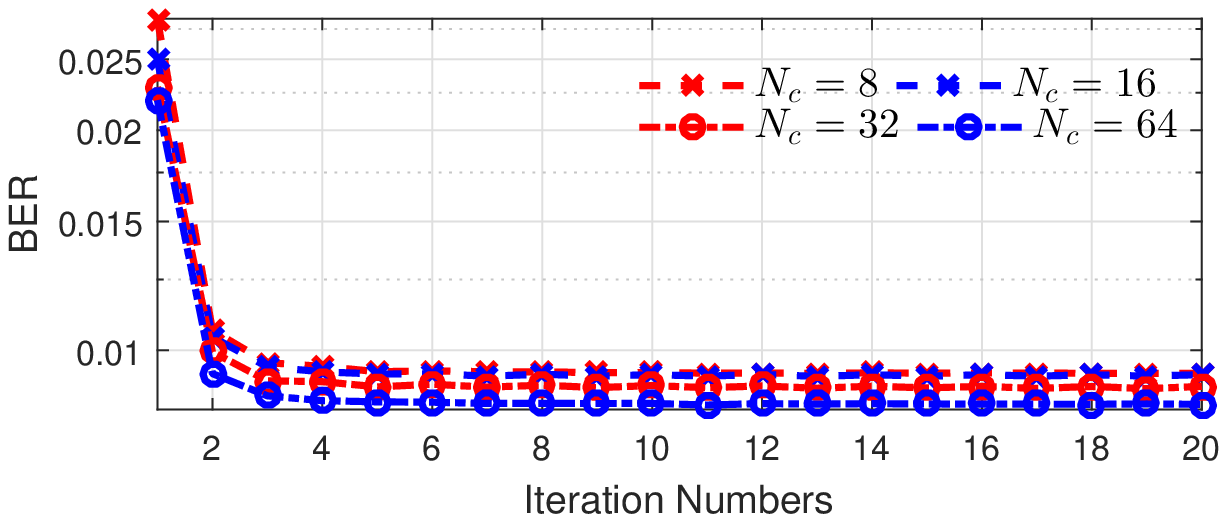}\label{F9B}}
	\caption{BERs versus iteration numbers of the proposed
		EP detector (namely, Algorithm 2) under different values of $N_c$ and SNR of -5 and 0 dB.\label{F9}}
\end{figure}

\begin{figure}[!t]
	\subfloat[BER vs $N_c$]{\includegraphics[scale=0.55]{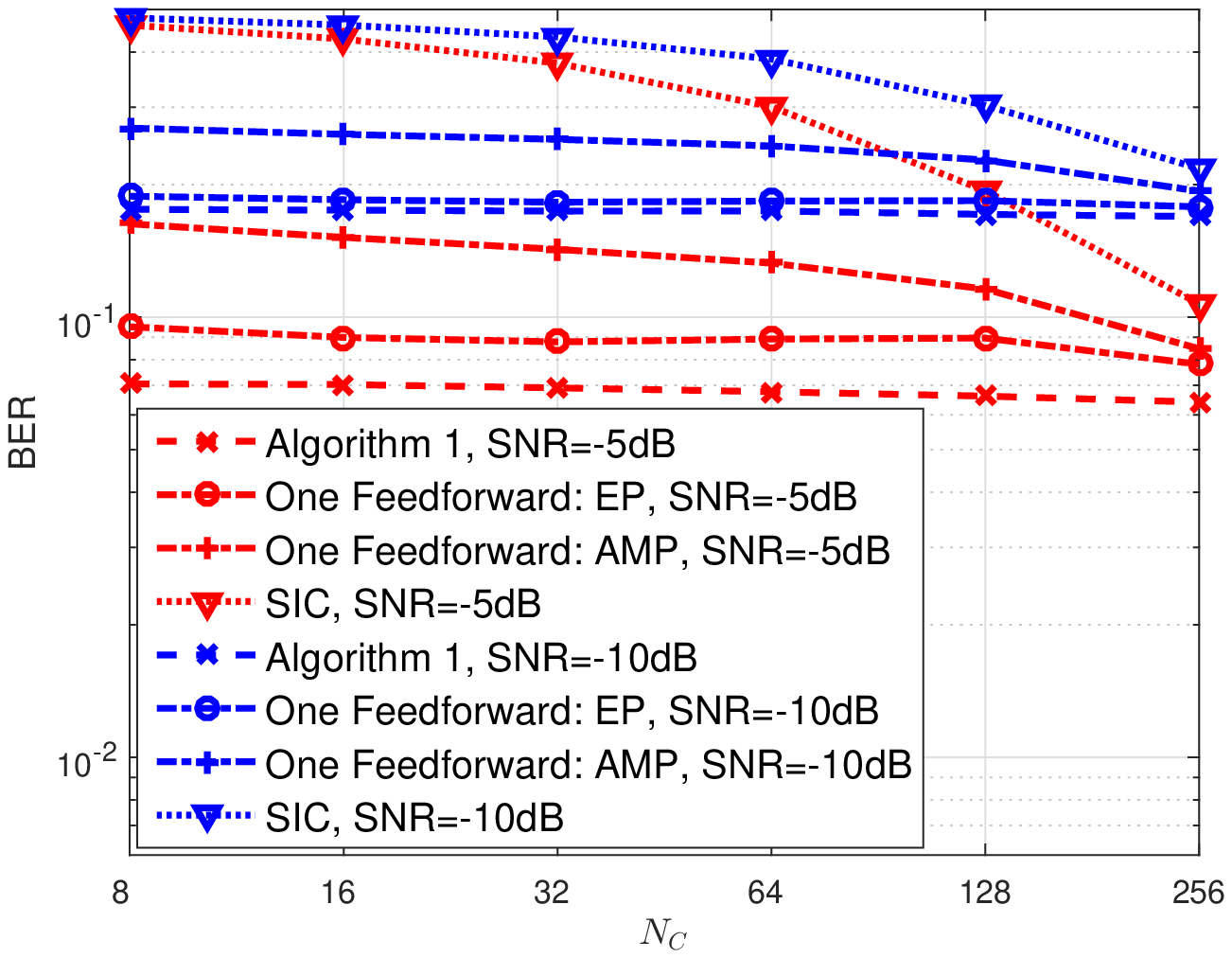} \label{F11A}}\vspace{-10pt}\\
	\subfloat[BER vs SNR]{\includegraphics[scale=0.55]{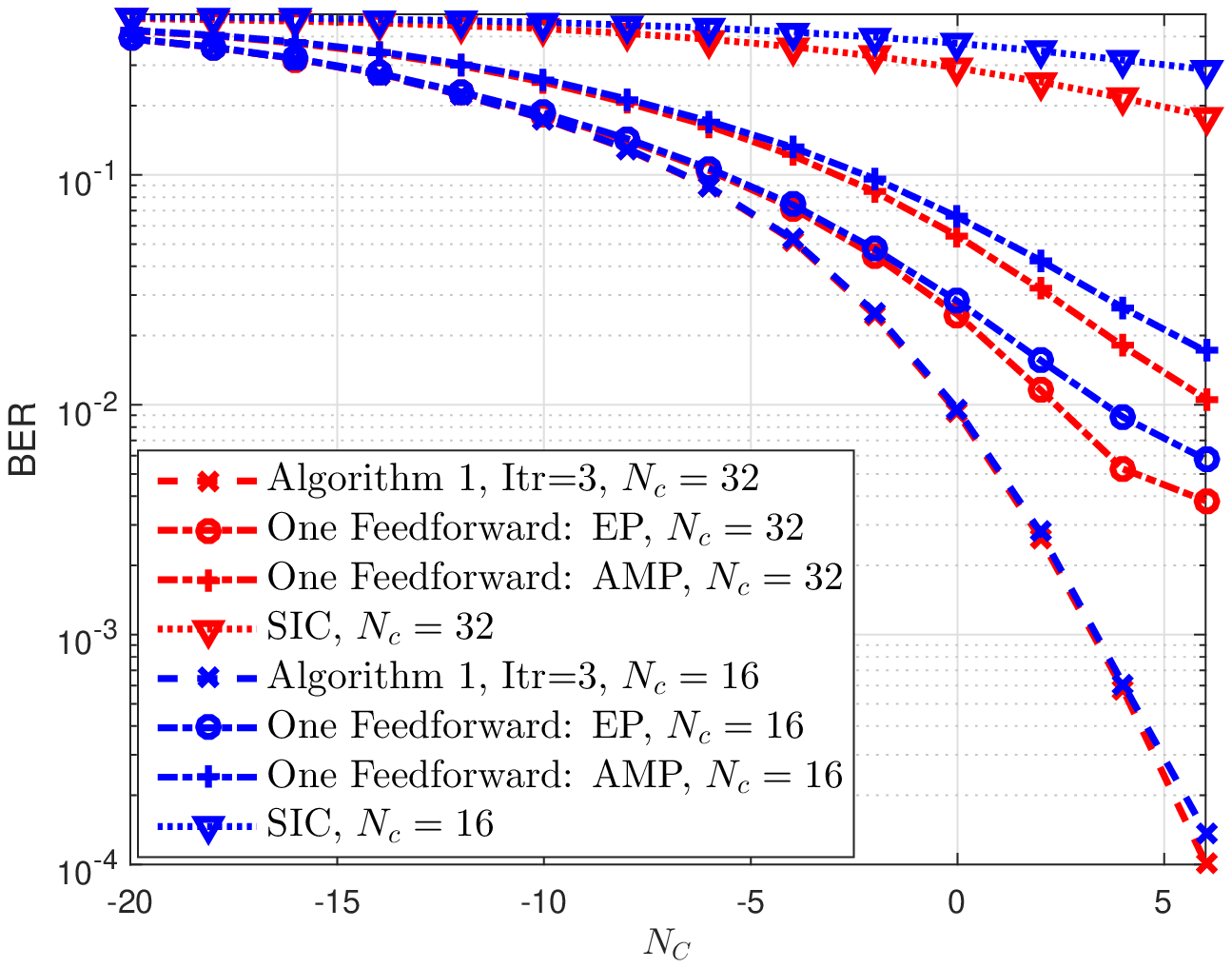} \label{F11B}}
	\caption{Comparison of the BER performances of various
		detectors considering the trimmed subarray model under different values of $N_c$ and SNR.\label{F11}}
\end{figure}

Fig. \ref{F9} shows the BER versus the iteration numbers for the EP-based detector derived for the trimmed subarray model (Section IV-A) under different $N_c$ and SNRs.
In all cases, the reductions in the BER values are minor after two iterations, meaning that  numerical stability is reached.
Thus, in the practical application, we recommend to perform two or three iterations.
In addition, as the value of $N_c$ decreases, slight performance loss can be observed (this finding can also be  observed in Fig. \ref{F11A}).
Hence, in practical scenario with non-i.i.d. channel matrix, implementing the hierarchical implementation architecture proposed in Section IV-B is only at a minor cost of BER performance.

As shown in Fig. \ref{F11}, we compare the BER performances of different algorithms that are proposed for the trimmed subarray model, including the EP-based detector derived for the trimmed subarray model (Section IV-A), the EP-based and approximate message passing (AMP)-based one feedforward architecture \footnote{These schemes refer to implement EP and AMP in parallel in each LPM by AMP in the one feedforward architecture presented in Section IV-C.}, and the SIC-based method proposed in \cite[Algorithm 3]{Amiri-2018GCW}.
Fig. \ref{F11A} shows the curve of BER versus $N_c$, and Fig. \ref{F11B} shows that of BER versus SNR.
In general, the EP-based algorithms outperform AMP- and SIC-based algorithms proposed in \cite[Algorithm 3]{Amiri-2018GCW}.
When applying EP-based algorithms, we prefer to set a smaller $N_c$ because doing so allows us to parallelize the matrix inversion to reduce complexity.
However, when applying AMP-based one feedforward method \cite[Algorithm 3]{Amiri-2018GCW}, we tend to set larger $N_c$ because setting smaller $N_c$ suffers from significant performance loss.
In addition, in the small SNR regime, the performance degradation of the modified EP-based detector requiring one feedforward is minor.
As SNR increases, this performance degradation further increases.
Therefore, the one feedforward method is favorable in the small SNR regime.

\section{Conclusion}
In this paper, the EP principle was exploited for the derivation of an efficient detector of  extra-large-scale massive MIMO systems with the subarray-based processing architecture.
We described the a posteriori distribution as a factor graph and developed the iterative algorithm by computing and transferring messages among different nodes on the factor graph.
We intuitively verified the convergence of the proposed algorithm via the evolution analysis.
We also characterized the fixed points of the algorithm.
We proved that these fixed points are the stationary points of the Bethe free energy optimization subject to the moment-matching constraints and are identical to the fixed point equations derived from the replica method.
We also discussed many implementation-related aspects, such as how to exploit non-stationarity for complexity reduction, the hierarchical implementation architecture to allow for parallel simple computation of matrix inversion, and the design of an algorithm with only one feedforward.
Finally, we demonstrated through simulation results that the proposed detector outperforms its counterparts and verified the validity of our analysis.

\appendices
\section{Message-Passing Derivation of Algorithm \ref{A1}}
In this appendix, we present the message-passing derivation of Algorithm \ref{A1}.
Following the EP principle \cite{Minka-2001,Minka-2005,Vehtari-2014Arxiv}, we constrain the messages to dwell in the family $\calF$ of Gaussian-distributed vectors with independent elements, which is in the form of $\calCN\left(\qx;\qm,\diag(\qv)\right)$.
For a factor graph with factor nodes $\{f_0,f_1,\ldots,f_C\}$, all connected to a set of variable node $\{\qx_i\}$, the messages are computed and updated iteratively by performing the steps below.

\begin{enumerate}
	\item \emph{Factor selection}: One or a set of factors is selected from $\{f_0,f_1,\ldots,f_C\}$.
	\item \emph{Variable-to-factor messages}: The message(s) from a variable node (e.g., $\qx_j$) to the selected factor node(s) (e.g., $f_i$) is computed by
	\begin{equation}\label{fac_to_var}
	{\mu _{\qx_j \to {f_i}}}\left( \qx_j \right) = \prod\limits_{i' \ne i} {{\mu _{{f_{i'}} \to \qx_j}}\left( \qx_j \right)}.
	\end{equation}
	\item \emph{Factor-to-variable messages}: To compute the message(s) from the selected factor node(s) (e.g., $f_i$) to a variable node (e.g., $\qx_j$), the approximate belief $b_{f_i}(\qx_j)$ is initially computed at $f_i$, which is given by
	\begin{equation}\label{var_to_fac}
	\begin{aligned}
	&b_{f_i}(\qx_j)\!=\!\mathop{\arg \min }\limits_{b(\qx_j)\in {\calF}}\!{D_1}\!\left[ {\mu_{{{\bf{x}}_j} \to {f_i}}}\left( {{{\bf{x}}_j}} \right)\!\int\!{{f_i}} \left( {{{\bf{x}}_j},{{\left\{ {{{\bf{x}}_{j'}}} \right\}}_{j' \ne j}}} \right)\right.\\
	&~~~~~~~~\times\left.\left.\prod\limits_{j' \ne j} {{\mu _{{{\bf{x}}_{j'}} \to {f_i}}}\left( {{{\bf{x}}_{j'}}} \right){\mathop{\rm d}\nolimits} {{\bf{x}}_{j'}}}  \right\|b({{\bf{x}}_j}) \right],
	\end{aligned}
	\end{equation}
	where the divergence measure for two distributions, such as $p(x)$ and $q(x)$, ${D_1}\left[ {p\left( x \right)\left\| {q\left( x \right)} \right.} \right]$ is defined by
	\[\begin{aligned}
	{D_1}\left[ {p\left( x \right)\left\| {q\left( x \right)} \right.} \right]&=\int {p\left( x \right)\log \frac{{p\left( x \right)}}{{q\left( x \right)}}} {\rm{d}}x\\
	&+\int{\left( {q\left( x \right)-p\left( x \right)} \right)} {\rm{d}}x,
	\end{aligned}\]
	Then, we set the message ${\mu _{{f_i} \to \qx_j}}\left( \qx_j \right)$ as ${\mu _{{f_i} \to \qx_j}}\left( \qx_j \right) \propto b_{f_i}(\qx_j)/{\mu_{\qx_j \to {f_i}}(\qx_j)}$.	
\end{enumerate}

\begin{Remark}
	The factor graph in Fig. \ref{F2A} has only one variable node $\qx$, then \eqref{var_to_fac} reduces to
	\begin{equation}\label{var_to_fac_sim}
	b_{f_i}(\qx)=\arg\min\limits_{b(\qx)\in\calF}D_{1}\left[ {{\mu _{\qx \to {f_i}}}\left( \qx \right){f_i}\left( \qx \right)}\|b(\qx) \right].
	\end{equation}
	Now, we consider the minimization in \eqref{var_to_fac_sim}.
	Following the result in \cite{Minka-2005}, the belief $b_{f_i}(\qx)\in\calF$ minimizing $D_1$ in \eqref{var_to_fac_sim} is obtained as $\calCN\left(\qx;\qm_q,\diag(\qv_q)\right)$, where $\qm_q$ and $\qv_q$ denote the vectors comprising the mean and variance of each element of $\qx$ w.r.t.
	the distribution ${{\mu _{\qx \to {f_i}}}\left( \qx \right){f_i}\left( \qx \right)}$ respectively.
	In addition, similar to strategies utilized in \cite{Yuan-2014TWC,Rangan-2016Arxiv,JMA-2016SPL,JMa-2016Access}, we take the average of $\qv_q$ as the variance for the belief, which yield $b_{f_i}(\qx)\propto\calCN\left(\qx;\qm_q,K^{-1}\tr(\diag(\qv_q))\qI_K\right)$, to save the computational resources when implementing the algorithms.	
\end{Remark}

Then, we elaborate how to construct Algorithm \ref{A1} from the aforementioned rules.
In accordance to the decentralized processing architecture in Fig. \ref{F1}, we alternate the associated Steps (2) and (3) between the selection of factors $\{f_1,\ldots,f_C\}$ and that of factor $f_0$, which corresponds to the processing in each LPM and that at the CPM.
At the first iteration, we skip the Step (2) and initialize the message from the variable node $\qx$ to factors $\{f_1,f_2,\ldots,f_C\}$ by ${\mu _{\qx \to {f_c}}}\left( \qx \right)=\calCN\left( \qx;\qzero,E_x\qI \right)$ for $c\in\calC$.
Notably, in Algorithm \ref{A1}, $\tau_{c}$ and $\qgamma_{c}$ denote the reciprocal of the variance and mean of messages $\{{\mu _{\qx \to {f_c}}}\}_{c=0}^{C}$ respectively, namely, ${\mu _{\qx \to {f_c}}}(\qx)=\calCN\left(\qx;\qgamma_{c},\tau^{-1}_{c}\qI\right)$.
Therefore, the initialization of Algorithm \ref{A1} along with \eqref{pri_var_dec} and \eqref{pri_m_dec} yield the above initializations for ${\mu _{\qx \to {f_c}}}\left( \qx \right)$'s.
Then, we proceed to Step (3), which starts by computing the approximate belief $\{b_{f_c}(\qx)\}_{c=0}^{C}$ for $c\in\calC$.
To solve the $D_1$-minimization problem in \eqref{fac_to_var}, we compute the a posteriori mean and covariance matrix w.r.t. the likelihood functions ${f_c}\left(\qx \right) = \calCN(\qy;\qH_c\qx,\sigma ^2\qI)$ and the Gaussian a priori ${\mu _{\qx \to {f_c}}}\left( \qx \right)$.
From \cite[Theorem 12.1]{Kay1993}, this a posteriori distribution is Gaussian with covariance matrix $\qSigma_c$ and mean $\hat{\qx}_c$ derived as \eqref{lmmse_var_dec} and \eqref{lmmse_mean_dec}, respectively.
Then we take the reciprocal of the average diagonal elements for $\qSigma_c$ as $\omega_c$ in \eqref{ave_mmse_dec}.
Finally, we have $b_{f_c}(\qx)=\calCN\left(\qx;\hat{\qx}_c,\omega^{-1}_c\qI\right)$.
Next, we set the message ${\mu _{{f_c} \to \qx}}\left( \qx \right)$ to be the ratio of the approximate belief $b_{f_c}(\qx)$ to the most recent message ${\mu _{\qx \to {f_c}}}\left( \qx \right)$, which is given by
\begin{equation}\label{Guas_div_c}
	\begin{aligned}
		{\mu _{{f_c} \to \qx}}\left( \qx \right)&=\frac{b_{f_c}(\qx)}{{\mu _{\qx \to {f_c}}}\left( \qx \right)}=\frac{\calCN\left(\qx;\hat{\qx}_c,\omega^{-1}_c\qI\right)}{\calCN\left(\qx;\qgamma_{c},\tau^{-1}_{c}\qI\right)}\\
		&\propto\calCN\left(\qx;\frac{\omega_{c}\hat{\qx}_c-\tau_{c}\qgamma_{c}}{\omega_{c} - \tau_{c}},(\omega_{c} - \tau_{c})^{-1}\qI\right).
	\end{aligned}
\end{equation}
Then, we have ${\mu _{{f_c} \to \qx}}\left( \qx \right)\propto\calCN\left(\qx;\rqx_c,\eta^{-1}_c\qI \right)$ with $\eta_c=\omega_{c} - \tau_{c}$ and $\rqx_c=\left(\omega_{c}\hat{\qx}_c-\tau_{c}\qgamma_{c}\right)/\eta_c$, which yields \eqref{ext_var_dec} and \eqref{ext_m_dec}.

Subsequently, we select the factor $f_0$.
In accordance with Step (2), we set the message ${\mu _{\qx \to {f_0}}}\left( \qx \right)$ to be the product of messages passing from factors $\{f_1,f_2,\ldots,f_C\}$ to $\qx$, then
\begin{equation}\label{Guas_mul_0}
	\begin{aligned}
		{\mu _{\qx \to {f_0}}}\left( \qx \right) &= \prod\limits_{c\in\calC} {{\mu _{{f_c} \to \qx}}\left( \qx \right)}=\prod\limits_{c\in\calC}\calCN\left(\qx;\rqx_c,\eta^{-1}_c\qI \right)\\
		&\propto\calCN\left(\qx;\frac{\sum_{c\in\calC}\eta_c\rqx_c}{\sum_{c\in\calC} \eta_c},\left(\sum_{c\in\calC} \eta_c\right)^{-1}\qI\right).
	\end{aligned}
\end{equation}
Then, we have ${\mu _{\qx \to {f_0}}}\left( \qx \right)\propto\calCN\left(\qx;\qgamma_{0},\tau^{-1}_0\qI \right)$ with $\tau_{0} = \sum_{c\in\calC}\eta_c$ and $\qgamma_{0} =\sum_{c\in\calC}\eta_c\rqx_c/\tau_{0}$, which yields \eqref{input_snr} and \eqref{input_obs}.
Following Step (3), we set the approximate belief on $f_0$ as $b_{f_0}(\qx)\propto\calCN\left(\qx;\hat{\qx}_0,\omega^{-1}_0\qI\right)$ with $\omega_0=\sum_{k\in\calK}v_{0,k}/K$, where $\hat{x}_{0,k}$ and $v_{0,k}$ denote the MMSE estimate and its corresponding MSE of $x_k$ from its AWGN observation modeled by \eqref{AWGN_obs}, whose explicit expression are given by \eqref{mmse_est_fus}.
Similar to \eqref{Guas_div_c}, we set the message ${\mu _{{f_0} \to \qx}}\left( \qx \right)$ as
\[\begin{aligned}
{\mu _{{f_0} \to \qx}}\left( \qx \right)&=\frac{b_{f_0}(\qx)}{{\mu _{\qx \to {f_0}}}\left( \qx \right)}=\frac{\calCN\left(\qx;\hat{\qx}_0,\omega^{-1}_0\qI\right)}{\calCN\left(\qx;\qgamma_{0},\tau^{-1}_{0}\qI\right)}\\
&\propto\calCN\left(\qx;\frac{\omega_{0}\hat{\qx}_c-\tau_{0}\qgamma_{0}}{\omega_{0} - \tau_{0}},(\omega_{0} - \tau_{0})^{-1}\qI\right).
\end{aligned}\]
Finally, we denote $\eta_0=\omega_{0} - \tau_{0}$ and $\lqx_0=\left(\omega_{0}\hat{\qx}_0-\tau_{0}\qgamma_{0}\right)/\eta_0$, and the current iteration is completed.
Afterwards, the next iteration begins, and the factors $\{f_1,\ldots,f_C\}$ are selected again.
From Step (2) and similar to \eqref{Guas_mul_0}, we compute the message ${\mu _{\qx \to {f_c}}}(\qx)$ by
\begin{equation}\label{Guas_mul_c}
\begin{aligned}
&{\mu _{\qx \to {f_c}}}\left( \qx \right) = {{\mu _{{f_0} \to \qx}}\left( \qx \right)}\prod\limits_{c'\in\calC, c'\neq c} {{\mu _{{f_c'} \to \qx}}\left( \qx \right)}\\
&\propto\calCN\left(\qx;\frac{\eta_0\lqx_0+\sum_{c'\in\calC, c'\neq c}\eta_{c'}\rqx_{c'}}{\eta_0+\sum_{c'\in\calC, c'\neq c} \eta_{c'}}\right.,\\
&~~~~\left.\left(\eta_0+\sum_{c'\in\calC, c'\neq c} \eta_{c'}\right)^{-1}\qI\right)=\calCN\left(\qx;\qgamma_{c},\tau^{-1}_{c}\qI\right).
\end{aligned}
\end{equation}
Then, we can simply the expression of $\tau_{c}$ as
\[\tau_{c}=\eta_0+\sum_{c'\in\calC, c'\neq c} \eta_{c'}=\omega_{0}-\tau_{0}+\tau_{0}-\eta_c=\omega_{0}-\eta_c,\]
which is identical to \eqref{pri_var_dec}. In the similar manner, we can simplify $\qgamma_{c}$ as \eqref{pri_m_dec}.
Repeating the above message update procedure yields Algorithm \ref{A1}.

\section{Proof of Proposition \ref{evo_converg}}
We start by defining the following auxiliary functions:
\begin{subequations} \label{eq:Aux_fun}
	\begin{align}
	&\Phi_c\left(\nu\right)=(\mse_c\left(\nu\right))^{-1}-\nu^{-1}~~\mathrm{for}~~c\in\calC,\displaybreak[0]\\
	&\Psi\left(\rho\right)=(\mse_0\left(\rho\right))^{-1}-\rho^{-1}.
	\end{align}		
\end{subequations}
Then, we can rewrite the evolution equations in \eqref{eq:SE1} and \eqref{eq:SE2} w.r.t. $\{\Phi_c\}_{c=0}^{C}$ and $\Psi$ as follows
  can be re written
  \vspace{-5pt}
\begin{subequations}
	\begin{align}
	&\rho^{t}=\left(\sum_{c=1}^C\Phi_c\left(\nu^t_c\right)\right)^{-1},\displaybreak[0]\\
	&\nu^{t+1}_c=\left(\Psi\left(\rho^{t}\right)+\sum_{c'\neq c}\Phi_{c'}\left(\nu^t_{c'}\right)\right)^{-1}~~\mathrm{for}~~c\in\calC,\label{eq:alter_SE2}
	\end{align}		
\end{subequations}
where \eqref{eq:alter_SE2} follows from the manipulation of \eqref{eq:SE2} shown as
\[\nu^{t+1}_c=\left(\frac{1}{\mse_0\left(\rho^{t}\right)}-\frac{1}{\rho^{t}}+\sum_{c'\neq c}\left(\frac{1}{\mse_{c'}\left(\nu^t_{c'}\right)}-\frac{1}{\nu^t_{c'}}\right)\right)^{-1}.\]
We learn from the appendices in \cite{JMa-2016Access} that $\Phi_c$'s and $\Psi$ are monotonically decreasing for nonnegative $\nu$ and $\rho$ and that
\vspace{-5pt}
\[\lim\limits_{\rho\to\infty}\Psi\left(\rho\right)=E^{-1}_{x}=(\nu^{1}_1)^{-1}=\cdots=(\nu^{1}_C)^{-1},\]
As $\Psi\left(\rho\right)$ is monotonically decreasing, we conclude that for $\rho>0$,
\begin{equation}\label{pos_Psi}
	\Psi\left(\rho\right)\geq\lim\limits_{\rho\to\infty}\Psi\left(\rho\right)=E^{-1}_{x},
\end{equation}
and evidently, $\Psi\left(\rho\right)>0$. We also notice that
\[\mse_c\left(\nu_c\right)=\frac{1}{N_c}\sum\limits_{i=1}^{N_c}\frac{\sigma^2\nu_c}{\lambda_{i,c}\nu_c+\sigma^2}=\frac{1}{N_c}\sum\limits_{i=1}^{N_c}\frac{1}{\frac{\lambda_{i,c}}{\sigma^2}+\frac{1}{\nu_c}}\leq\nu_c\]
recalling that $\lambda_{i,c}\geq0$, which follows directly that $\Phi_c\left(\nu\right)\geq 0$ for $\nu\geq 0$ and $\forall c\in\calC$.

On the basis of the above results, we prove the monotonicity of the sequences $\{\rho^{t}\}$ and $\{\nu^{t}_c\}_{c=1}^{C}$ by induction.
The monotonicity of $\{\nu^{t}_c\}$ can be proved by induction.
Initially, we have
\[\nu^{2}_c=\left(\Psi\left(\rho^{1}\right)+\sum_{c'\neq c}\Phi_{c'}\left(\nu^1_{c'}\right)\right)^{-1}>E_{x}=\nu^{1}_c,\]
where the inequality follows from \eqref{pos_Psi} and the non-negativity of $\Phi_c$'s.
Then, we proceed by supposing that $\nu^{t}_c<\nu^{t-1}_c$.
As $\{\Phi_c\}_{c=1}^{C}$ are monotonically decreasing, we have $\Phi_c\left(\nu^{t}_c\right)>\Phi_c\left(\nu^{t-1}_c\right)$, which follows directly that
\begin{equation}\label{mono_rho}
	\rho^{t}=\left(\sum_{c=1}^C\Phi_c\left(\nu^t_c\right)\right)^{-1}<\left(\sum_{c=1}^C\Phi_c\left(\nu^{t-1}_c\right)\right)^{-1}=\rho^{t-1}.	
\end{equation}
Then from the monotonicity of $\Psi$, we have $\Psi\left(\rho^{t}\right)>\Psi\left(\rho^{t-1}\right)$. Integrating the fact that $\Phi_c\left(\nu^{t}_c\right)>\Phi_c\left(\nu^{t-1}_c\right)$, we find that
\[\begin{aligned}
\nu^{t+1}_c&=\left(\Psi\left(\rho^{t}\right)+\sum_{c'\neq c}\Phi_{c'}\left(\nu^t_{c'}\right)\right)^{-1}\\
&<\left(\Psi\left(\rho^{t-1}\right)+\sum_{c'\neq c}\Phi_{c'}\left(\nu^{t-1}_{c'}\right)\right)^{-1}=\nu^{t}_c.
\end{aligned}\]
Hence, $\{\nu^{t}_c\}_{c=1}^{C}$ are monotonically decreasing sequences.
In addition, the monotonicity of $\{\rho^{t}\}$ can directly follow from \eqref{mono_rho}.

For their boundedness, the non-negativity of $\Phi_c$'s and $\Psi$ implies that the sequences $\{\rho^{t}\}$ and $\{\nu^{t}_c\}$'s have the lower bound $0$.
Meanwhile, their monotonicity implies they have upper bound $\nu^{1}_c$ and $\rho^{0}$, respectively.

\section{Proof of Proposition \ref{free_ener_min}}
In this appendix, we prove Proposition \ref{free_ener_min} via optimality condition for the equality constrained problem \cite[Sec. 3.1]{Bertsekas1995}.
We define the Lagrangian function corresponding to the optimization problem in \eqref{BFE_Min} as
\[\begin{aligned}
&L(b_0,b_1,\ldots,b_C,q,\qalpha,\qbeta,\qmu)=J\left(b_0,b_1,\ldots,b_C,q\right)\\
&+\sum\limits_{c=0}^{C}\frac{\mu_c}{K}\left(\sum\limits_{k\in\calK}\Ex_{b_c}(|x_k|^2)-\sum\limits_{k\in\calK}\Ex_{q}(|x_k|^2)\right)\\
&+\sum\limits_{c=0}^{C}\qalpha^{T}_c\Re\left(\Ex_{b_c}(\qx)-\Ex_{q}(\qx)\right)+\sum\limits_{c=0}^{C}\qbeta^{T}_c\Im\left(\Ex_{b_c}(\qx)-\Ex_{q}(\qx)\right).
\end{aligned}\]
To show that $b_0(\qx),b_1(\qx),\ldots,b_C(\qx)$ and $q(\qx)$ in \eqref{stat_points} are the stationary points of the problem \eqref{BFE_Min}, we show that there exists Lagrange multipliers $\qalpha_0,\qalpha_1,\ldots,\qalpha_C,\qbeta_0,\qbeta_1,\ldots,\qbeta_C$ and $\qmu=[\mu_0,\mu_1,\ldots,\mu_C]^T$ such that $b_0(\qx),b_1(\qx),\ldots,b_C(\qx)$ and $q(\qx)$ in \eqref{stat_points} satisfy
\begin{subequations}
	\begin{align}
		b_c&=\arg\min\limits_{b_c}~L(b_0,b_1,\ldots,b_C,q,\qalpha,\qbeta,\qmu)~\text{for}~c=0,1,\ldots,C,\label{min_1}\\
		q&=\arg\min\limits_{q}~L(b_0,b_1,\ldots,b_C,q,\qalpha,\qbeta,\qmu).\label{min_2}
	\end{align}
\end{subequations}
Let $\qalpha_c=-\Re(\tau_{c}\qgamma_{c})$, $\qbeta_c=-\Im(\tau_{c}\qgamma_{c})$, and $\mu_c=K\tau_{c}$, we can write the Lagrangian function w.r.t. $b_c$ as follows
\[\begin{aligned}
&L(b_0,b_1,\ldots,b_C,q,\qalpha,\qbeta,\qmu)\\
&=\!\KL\left(b_c\|e^{\log f_c(\qx)}\right)\!\!-\!\!2\Re\left(\tau_{c}\qgamma^{H}_{c}\Ex_{b_c}(\qx)\right)\!\!+\!\!\tau_{c}\Ex_{b_c}(\|\qx\|^2)\!\!+\!\!\text{const}\\
&=-H(b_c)+\Ex_{b_c}\left[\log f_c(\qx)+\tau_{c}\|\qx-\qgamma_{c}\|^2\right]+\text{const}\\
&=\KL\left(b_c\left\|\frac{1}{Z_c\left(\qgamma_{c}\right)}\exp\left[\log f_c(\qx)-\tau_c\|\qx-\qgamma_{c}\|^2\right]\right)\right.+\text{const},
\end{aligned}\]
where the constant terms are independent on $b_c$, and the last two equalities follow directly from the definition of KL divergence.
Therefore, $b_0(\qx),b_1(\qx),\ldots,b_C(\qx)$ defined in \eqref{stat_points} achieves the minimization in \eqref{min_1}.

Similarly, we can obtain the maximization in \eqref{min_2}by rewriting the Lagrangian function w.r.t. $q$ as follows
\[\begin{aligned}
&L(b_0,b_1,\ldots,b_C,q,\qalpha,\qbeta,\qmu)\\
&\stackrel{(*)}{=}C\left\{H(q)+2\Re\left(\omega\hat{\qx}^{H}\Ex_{q}(\qx)\right)-\omega\Ex_{q}(\|\qx\|^2)\right\}+\text{const}\\
&=C\left\{H(q)+\Ex_{q}\left[\omega\|\qx-\hat{\qx}\|^2\right]\right\}+\text{const}\\
&=-C\KL\left(q\left\|\frac{1}{Z_q\left(\hat{\qx}\right)}\exp\left[-\omega\|\qx-\hat{\qx}\|^2\right]\right)\right.+\text{const},
\end{aligned}\]
where $(*)$ follows from Lemma \ref{Fixed_Point} that $\sum_{c=0}^{C}\tau_{c}=C\omega$ and $\sum_{c=0}^{C}\tau_{c}\qgamma_{c}=C\omega\hat{\qx}$.
Hence, the maximization in \eqref{min_2} can be attained by $q(\qx)$ defined in \eqref{stat_points}.

In addition, the moment-matching constraints in \eqref{Moment_Matching} can be justified to satisfy by noting that $\hat{\qx}$ and $K\omega^{-1}$ are just the mean and trace of covariance matrices of the densities $b_0(\qx),b_1(\qx),\ldots,b_C(\qx)$ from \eqref{lmmse_mean_dec}, \eqref{ave_mmse_dec}, \eqref{post_mean_fus}, and \eqref{post_var_fus} and that $q(\qx)$ is Gaussian with mean $\qx$ and covariance matrix $w^{-1}\qI_K$.

\section{Proof of Proposition \ref{fixed_point_equ}}
From the definition of the Stieltjes transform \cite[Sec. 2.2.1]{Tulino-2004}, when $K\to\infty$,
\[\omega^{-1}_c = \frac{1}{K}\tr\left({\left(\sigma^{-2} \qH_c^H\qH_c+ \tau_{c}\qI\right)}^{-1}\right)\to S_{\sigma^{-2}\qH_c^H\qH_c}(-\tau_{c}).\]
Then, we have $\tau_{c}=-S^{-1}_{\sigma^{-2}\qH_c^H\qH_c}(\omega^{-1}_c)$, where $S^{-1}_{\sigma^{-2}\qH_c^H\qH_c}(\cdot)$ denotes the inverse function of $S_{\sigma^{-2}\qH_c^H\qH_c}(\cdot)$. Hence, from \eqref{ext_var_dec} and the definition of R-transform \cite[Sec. 2.2.5]{Tulino-2004}, we have
\[\eta_c = \omega_{c} - \tau_{c}=\omega_{c}+S^{-1}_{\sigma^{-2}\qH_c^H\qH_c}(\omega^{-1}_c)=R_{\sigma^{-2}\qH_c^H\qH_c}(-\omega^{-1}_c).\]
According to Lemma \ref{Fixed_Point}, for any fixed point, $R_{\sigma^{-2}\qH_c^H\qH_c}(-\omega^{-1}_c)=R_{\sigma^{-2}\qH_c^H\qH_c}(-\omega^{-1})$ for $\forall c\in\calC$.
We use $\hat{\qh}_{c,i}$ and $\hat{\qh}_{i}$ to denote the $i$-th column of the matrix $\qH^H_{c}$ and $\qH^H$ respectively.
Then, following the argument in \cite[Sec. 2.4.2]{Tulino-2004} and from \eqref{input_snr}, we can prove that when $\qH$ is a random matrix with i.i.d. elements,
\[\begin{aligned}
	\tau_{0} &= \sum_{c=1}^C \eta_c=\sum_{c=1}^CR_{\sigma^{-2}\qH_c^H\qH_c}(-\omega^{-1})\\
	&=\sum_{c=1}^C\left(\lim\limits_{N_c\to\infty}\sum_{i=1}^{N_c}R_{\sigma^{-2}\hat{\qh}_{c,i}\hat{\qh}^H_{c,i}}(-\omega^{-1})\right)\\
	&=\lim\limits_{N\to\infty}\sum_{i=1}^{N}R_{\sigma^{-2}\hat{\qh}_{i}\hat{\qh}^H_{i}}(-\omega^{-1})=R_{\sigma^{-2}\qH^H\qH}(-\omega^{-1}).
\end{aligned}\]
The above arguments are derived from the asymptotic freeness of the relevant matrices, which also requires $K\to\infty$.
As such, we complete the proof of \eqref{Replica_Result}.

%
%

\ifCLASSOPTIONcaptionsoff
  \newpage
\fi



%

\bibliography{IEEEabrv,myBib}

%

\end{document}